\documentclass[fleqn,usenatbib]{mnras}

\usepackage{newtxtext,newtxmath}

\usepackage[T1]{fontenc}

\DeclareRobustCommand{\VAN}[3]{#2}
\let\VANthebibliography\thebibliography
\def\thebibliography{\DeclareRobustCommand{\VAN}[3]{##3}\VANthebibliography}

\usepackage{graphicx}	
\usepackage{amsmath}	
\usepackage[dvipsnames]{xcolor} 

\title[GWs from MHD supernova simulations]{The gravitational-wave emission from the explosion of a 15 solar mass star with rotation and magnetic fields}

\author[Powell \& M\"uller]{
Jade Powell,$^{1,2}$\thanks{E-mail: dr.jade.powell@gmail.com}
Bernhard M\"uller,$^{3}$
\\
$^{1}$ Centre for Astrophysics and Supercomputing, Swinburne University of Technology, Hawthorn, VIC 3122, Australia.\\
$^{2}$ ARC Centre of Excellence for Gravitational Wave Discovery (OzGrav), Melbourne, Australia. \\
$^{3}$ School of Physics and Astronomy, Monash University, VIC 3800, Australia. 
}

\pubyear{2024}

\begin{document}
\label{firstpage}
\pagerange{\pageref{firstpage}--\pageref{lastpage}}
\maketitle

\begin{abstract}
Gravitational waveform predictions from 3D simulations of explosions of non-rotating massive stars with no magnetic fields have been extensively studied. However, the impact of magnetic fields and rotation on the core-collapse supernova gravitational-wave signal is not well understood beyond the core-bounce phase. Therefore, we perform four magnetohydrodynamical simulations of the explosion of a $15\,M_{\odot}$ star with the SFHx and SFHo equations of state. All of the models start with a weak magnetic field strength of $10^{8}$\,G, and two of the models are rapidly rotating. We discuss the impact of the rotation and magnetic fields on the gravitational-wave signals. We find that the weak pre-collapse fields do not have a significant impact on the gravitational-wave signal amplitude. With rapid rotation, the f/g-mode trajectory can change in shape, and the dominant emission band becomes broader. We include the low-frequency memory component of the gravitational-wave signal from both matter motions and neutrino emission anisotropy. We show that including the gravitational waves from anisotropic neutrino emission increases the supernova detection distances for the Einstein Telescope, and would also be detectable out to Mpc distances by a moon-based gravitational-wave detector.  
\end{abstract}



\section{Introduction}

Core-collapse supernovae (CCSNe) are the explosive deaths of massive stars. Once the star's iron core reaches its effective Chandrasekhar mass, the core collapses until it reaches and overshoots nuclear saturation density. Due to the stiffening to the nuclear equation of state (EoS), the collapse is then halted, and a shock wave is launched outwards due to the rebound of the iron core. Due to energy losses by nuclear dissociation and neutrino losses, the shock quickly stalls, but still expands as an accretion shock to a radius of $\sim 150$\,km. Shock revival for the majority of CCSNe is expected to be powered by neutrinos,  which need to be sufficiently strong to deliver
explosion energies of $\sim 10^{51}$\,erg as
in observed CCSNe. Explosions powered by the
neutrino-driven mechanism have been simulated extensively to understand the dynamics, remnant properties, neutrinos, and the gravitational-wave (GW) emission of CCSNe (see \citet{mueller_20, 2020arXiv201004356A, burows_21} for recent reviews). However, the majority of 3D simulations of neutrino-driven explosions do not include the effects of rotation and magnetic fields. 

Magnetic fields and rotation are thought to play a critical role in hypernova explosions, which can reach much higher explosion energies of $\sim 10^{52}$\,erg \citep{woosley_06,iwamoto_98,mueller_24}. These explosions, which sometimes also produce gamma-ray bursts, are likely produced by a magnetorotational mechanism that taps the energy of a rapidly rotating milli-second magnetar through the magnetic field \citep{1992Natur.357..472U,duncan_92}, or a black hole accretion disk in collapsars \citep{1999ApJ...524..262M}. 
In recent years, significant advances have been made in 3D simulations of magnetorotational explosions \citep{2020ApJ...896..102K,  10.1093/mnras/staa3273, 2021MNRAS.tmp...72R, reichert_22, 2022MNRAS.512.2489O, 2023MNRAS.520.5622B, 2023MNRAS.522.6070P, shibagaki_24}. Long-duration simulations have shown the emergence of powerful jets, including some with kink instabilities \citep{moesta_14b}, and the ejection of the neutron rich  material needed for neucleosynthesis \citep{2012ApJ...750L..22W}. However, most simulations have failed to reach the explosion energies of $\sim 10^{52}$\,erg required to power a full hypernova explosion. One exception is \citet{obergaulinger_21}, who were able to reach an explosion energy of $\sim 10^{52}$\,erg for a model with an initial magnetic field strength of $10^{12}$\,G.  

The magnetic fields may not only play an important role in hypernovae, but also in more typical supernovae. Recent work has shown that neutrino-driven convection can drive a small scale turbulent dynamo behind the CCSN shock even when there is no rapid progenitor rotation \citep{2020MNRAS.498L.109M, 2022MNRAS.516.1752M, 2024MNRAS.528L..96M}. In particular, magnetic fields can become dynamically important \citep{2023MNRAS.526.5249V}
in the explosions of  CCSN progenitors with highly magnetised, but slowly rotating cores that originate from stellar mergers, which could be the progenitors of Galactic magnetars \citep{2019Natur.574..211S}. More magnetohydrodynamic simulations of regular CCSNe are required to understand the magnetic fields and spins of Galactic pulsars and magnetars, and the possible impact of the magnetic field on the explosion mechanism. 

CCSNe are a promising multi-messenger source for the current ground-based GW detectors Advanced LIGO \citep{2015CQGra..32g4001L}, Advanced Virgo \citep{2015CQGra..32b4001A} and KAGRA \citep{2021PTEP.2021eA102A}. No GWs have been detected from CCSNe in previous observing runs yet \citep[e.g.,][]{2020PhRvD.101h4002A, 2023arXiv230516146S}. Significant progress has been made in recent years predicting the GW emission for neutrino-driven CCSNe \citep{2019MNRAS.487.1178P, burrows_20, 2018ApJ...861...10M, 2017MNRAS.468.2032A, 2017ApJ...851...62K, 2020PhRvD.102b3027M, 2021ApJ...914..140P, 2019ApJ...876L...9R}. Although simulations from different groups can differ in their predictions of the GW amplitudes and precise frequency trajectories,
all groups have found the dominant feature of the GW signal to be the high frequency g/f-mode. Our understanding of the GW emission modes from 3D simulations has enabled several groups to develop phenomenological models to aid CCSN GW parameter estimation and searches \citep{2018PhRvD..98l2002A, 2022PhRvD.105f3018P, 2023arXiv230110019B}. 

The GW signal from CCSNe with rotation and magnetic fields is not so well understood. Energetic explosions with rapid rotation may have enough GW energy to be detected outside the Milky Way \citep{Szczepanczyk_2021}. Rotation results in a spike in the time-series at the core-bounce time, which has been extensively studied \citep{2002A&A...393..523D, 2017PhRvD..95f3019R, 2014PhRvD..90d4001A}, including the early post-bounce phase in 3D \citep{Scheidegger2008}. However, GW emission from the later phases of the explosion has not yet been investigated as thoroughly. Several studies have included rotation in their neutrino-driven explosions \citep{2018MNRAS.475L..91T, andresen_19, 2020MNRAS.494.4665P, shibagaki_20, 2021ApJ...914..140P, 2021MNRAS.508..966T}, and a few GW signals are now available for long-duration simulations that include both magnetic fields and rotation, where the magnetic fields played a major role in driving the explosion \citep{Jardine_2021, 2023MNRAS.520.5622B, 2023MNRAS.522.6070P}. 
The f/g-mode is also found to be the dominant feature in the GW emission in simulations that include rotation and magnetic fields. However, as the effects of magnetic fields and rotation have so far been neglected in CCSN astroseismology studies, it is less clear how the GW frequency relates to the proto-neutron star (PNS) properties. One noteworthy exception is the work of \citet{2021arXiv210312445R}, which addressed later phases beyond the first $\mathord{\sim}1\,\mathrm{s}$ of evolution, and
identified a clear imprint of magnetic fields from an $\alpha-\Omega$-dynamo in the PNS convection zone on GWs from inertial modes.

The standing accretion shock instability (SASI) \citep{0004-637X-584-2-971, 2006ApJ...642..401B, 2007ApJ...654.1006F} can also result in GW emission in the most sensitive frequency band of current GW detectors, and is often present in waveforms from neutrino-driven explosions. The GW amplitude of the SASI-driven modes is usually lower, and more difficult to detect, than the dominant g/f-mode. However, the majority of previous works investigating SASI signatures in the GW signal have not included both magnetic fields and rotation. CCSN simulations without neutrino transport have shown that SASI-driven turbulence can amplify magnetic fields exponentially \citep{2012ApJ...751...26E}. Due to strong shear flows at the neutron star surface, magnetic fields of order $\gtrsim 10^{14}$\,G may be achievable. 
However, the consequences of SASI-driven field amplification have not yet been investigated in self-consistent 3D CCSN simulations with neutrino transport, and the effects on the GW emission have not yet been investigated. 

Another important feature of the GW signals from CCSNe is the GW memory. 
General relativity predicts that matter or radiation from an asymmetric source should result in a permanent deformation of the space-time metric, known as GW memory. In supernova explosions, 
such a memory signal can arise from the asymmetric emission of neutrinos and from strongly asymmetric shock expansion \citep{1978ApJ...223.1037E, 2009ApJ...707.1173M, 2012A&A...537A..63M},
and results in GW emission below the frequency band of current GW detectors. Exactly how low
the spectrum of the memory signal reaches down in frequency is difficult to determine, as the minimum frequency that we can predict is equal to the inverse of the simulation time, and long duration simulations in 3D are too computationally expensive. However, many recent 3D studies have included the GW memory from matter, and a few 3D studies have started to also include the low frequency GW memory from asymmetric neutrino emission \citep{2023PhRvD.107j3015V}. The neutrino memory signal is expected to be a promising source for moon-based GW detectors \citep{2020arXiv200708550J, 2021ApJ...910....1H, 2024arXiv240409181A} and space based GW detectors such as DECIGO \citep{2021PTEP.2021eA105K} and LISA \citep{2017arXiv170200786A}. Some authors have produced analytical models to estimate the detectability of the CCSN low frequency signal in space detectors, with some studies estimating that these signals can be detected out to distances of 10\,Mpc, which would result in much higher rates of CCSNe in space detectors than in current ground-based interferometers \citep{ 2021arXiv210505862M, 2022PhRvD.105j3008R, 2024arXiv240513211G}.  

To further understand the GW signal from CCSNe in the presence of magnetic fields and rotation, we perform four magnetohydrodynamical simulations of a $15\,M_{\odot}$ progenitor star with two different EoS. All of the models start with a weak magnetic field strength of $10^{8}$\,G in the progenitor, and two of the models are rapidly rotating. We briefly discuss the hydrodynamics of the explosion and the PNS properties to aid in our description of the features of the GW emission. All of the models undergo rapid shock revival. 
We find that the magnetic fields do not significantly impact the GW amplitudes, which are $\sim 10$\,cm in our non-rotating models, and $\sim 40$\,cm in our rapidly rotating models. We show how the amplitude increases to as much as $\sim 200$\,cm when the component of the GW emission due to asymmetric neutrino emission is included. We show how this improves the maximum detectable distances for CCSN sources, using multiple different source angles. The low-frequency component of the GW signal from neutrino memory
is therefore an important aspect for the CCSNe science case for next-generation GW detectors. Some previous works have 
questioned whether universal relations for the g/f-mode frequency 
are still a good fit to the actual gravitational spectrograms predicted  by magnetorotational CCSN explosion models \citep{2023MNRAS.522.6070P}. We show in this work that our models are still a good fit for the universal relations when the magnetic fields are weak. 

Our paper is structured as follows: In Section~\ref{sec:sim} we describe the progenitor star and the setup of our simulations. 
Although the focus in this work is on the GW emission,
we give a brief overview of the explosion dynamics and the remnant properties
in Section~\ref{sec:dynamics}. In Section~\ref{sec:grav_waves_time}, we show the GW signals produced in our simulations. In Section~\ref{sec:neutrino}, we show the low frequency GW emission caused by neutrino emission anisotropy. The directional dependence of the
GW emission 
and the prospects for the detection of the
GW signals are investigated in Section~\ref{sec:direction}. In Section~\ref{sec:universal}, we analyse whether the magnetic fields and rotation impact the models fit to several universal relations proposed by different groups. A discussion and conclusions follow in Section~\ref{sec:conclusion}.

\section{Progenitor model and simulation methodology}
\label{sec:sim} 
The progenitor model for our simulations is the $15\,M_{\odot}$ model from \citet{1995ApJS..101..181W}, with some of the details for each of our four simulations provided in Table \ref{tab:models}. We use two different EoS, SFHo and SFHx from \citet{2013ApJ...774...17S}.
All four simulations have a dipolar initial magnetic field strength of $10^{8}$\,G both for the toroidal and poloidal field at the centre of the star. 
In two of the models we imposed rapid rotation by hand using the rotation law 
\begin{equation}
     \Omega =  \frac{\Omega_0}{
    (1+(r \sin \theta/r_0)^2},
\end{equation}
for the angular velocity $\Omega$,
with $\Omega_0=1 \, \mathrm{rad}\, \mathrm{s}^{-1}$
and $r_0=10^8 \,\mathrm{cm}$.
For the rest of this paper, we refer to the rapidly rotating model with the SFHo EoS as model SFHo\_rr, the rapidly rotating model with the SFHx EoS as model SFHx\_rr, the non-rotating model with the SFHo EoS as model SFHo\_nr, and the non-rotating model with the SFHx EoS as model SFHx\_nr. 
The post-bounce durations of the simulations are 0.42\,s for model SFHo\_rr, 0.48\,s for model SFHo\_nr, 0.41\,s for model SFHx\_rr, and 0.45\,s for model SFHx\_nr.

\begin{table*}
\centering
 \begin{tabular}{||c c c c c c c c ||} 
 \hline
  Model & Equation  & Rotating & Initial & Simulation & Explosion & PNS  & GW   \\ 
  name & of state &  & B field  & end time & energy & mass &  energy  \\
 \hline\hline
  SFHo\_rr & SFHo & Y & $10^8$\,G & 0.42\,s & $1.0\times10^{51}$\,erg & 1.34\,$M_{\odot}$ &  $4.2\times10^{47}$\,erg   \\
 \hline
  SFHo\_nr & SFHo & N & $10^8$\,G & 0.48\,s & $0.3\times10^{51}$\,erg & 1.48\,$M_{\odot}$ & $1.9\times10^{46}$\,erg   \\ 
 \hline
  SFHx\_rr & SFHx & Y & $10^8$\,G & 0.41\,s & $1.3\times10^{51}$\,erg & 1.35\,$M_{\odot}$ & $4.0\times10^{47}$\,erg    \\
 \hline
  SFHx\_nr & SFHx & N & $10^8$\,G & 0.45\,s & $0.2\times10^{51}$\,erg & 1.47\,$M_{\odot}$ & $1.6\times10^{46}$\,erg    \\
 \hline
\end{tabular}
\caption{For each model we list the model name, the equation of state, the initial core rotation rate, the initial magnetic field strength, the diagnostic explosion energies, the proto-neutron star mass at the end of the simulation time, and the GW energy at the end of the simulation time. Rotating models reach higher explosion energies and have stronger GW signals. Rotation also results in the formation of less massive neutron stars due to the earlier onset of the explosion. }
\label{tab:models}
\end{table*}

We perform our simulations using the magnetohydrodynamic (MHD) version of the \textsc{CoCoNuT-FMT} code as described in \citet{2020MNRAS.498L.109M}. The code solves the Newtonian MHD equations using the HLLC solver \citep{Gurski_2004, 2005JCoPh.208..315M} and hyperbolic divergence cleaning \citep{2002JCoPh.175..645D}. 
The  MHD equations 
for the density $\rho$, magnetic field $\mathbf{B}$, total energy density $e$, velocity $\mathbf{v}$ and Lagrangian multiplier $\psi$ are expressed in Equations~\eqref{eq:cont}--\eqref{eq:divclean}
in Gaussian units including divergence
cleaning terms as
\begin{align}
\partial_{t} \rho+\nabla \cdot(\rho \mathbf{v}) &=0,   \label{eq:cont}\\
\label{eq:mhd2}
\partial_{t}(\rho \mathbf{v})+\nabla \cdot\left[\rho \mathbf{v} \mathbf{v}^{\mathrm{T}}+\left(P+\frac{B^{2}}{8\pi}\right) \mathcal{I}-\frac{\mathbf{B} \mathbf{B}^{\mathrm{T}}}{4\pi}\right] &=\rho \mathbf{g}+\mathbf{Q}_\mathrm{m}
-\frac{(\nabla \cdot \mathbf{B}) \mathbf{B}}{4\pi}, \\
\partial_{t} \mathbf{B}+\nabla \cdot\left(\mathbf{v} \mathbf{B}^{\mathrm{T}}-\mathbf{B} \mathbf{v}^{\mathrm{T}}+c_\mathrm{h}\psi \mathcal{I}\right) &=0,  \\
\label{eq:mhd4}
\partial_{t} \hat{e}+\nabla \cdot\left[\left(e+P+\frac{B^2}{8\pi} \right) \mathbf{v}-\mathbf{B}(\mathbf{v} \cdot \mathbf{B})\right] &=
\rho \mathbf{v}\cdot \mathbf{g}+Q_\mathrm{e}
+\mathbf{Q}_\mathrm{m}\cdot\mathbf{v},
\\
\partial_{t} \psi+c_\mathrm{h} \nabla \cdot \mathbf{B} &=-\frac{\psi}{\tau} . 
\label{eq:divclean}
\end{align}
Here $\rho$, $\mathbf{v}$, $P$, and $e$, and $\mathbf{B}$ are the standard magnetohydrodynamic variables
density, velocity, pressure, total gas energy density and magnetic field. Among the other variables, $c_\mathrm{h}$ denotes the hyperbolic cleaning speed, $\tau$ the damping time for
the Lagrangian multiplier, and $Q_\mathrm{e}$ and $\mathbf{Q}_\mathrm{m}$ are the neutrino energy
and momentum source terms, and $\hat{e}
=e+(B^2+\psi^2)/(8\pi)+$ is the total energy density of the fluid, the magnetic field and the cleaning field.
The cleaning speed $c_\mathrm{h}$ is identified with the fast magnetosonic velocity, and the damping
time is set to eight times the magnetosonic crossing time of a cell.
The effective potential of \citet{mueller_08} is used to approximate
the effects of relativistic gravity.
The GW emission is extracted by the time-integrated quadrupole formula \citep{finn_89, finn_90, blanchet_90}. 
GW amplitudes are given as distance-normalized quadrupole
amplitudes \citep{thorne_80} $h R$ (strain $h$ times distance $R$), except where the distance is explicitly specified.
Neutrinos are treated using the \textsc{FMT} (Fast Multi-group Transport) method of \citet{2015MNRAS.448.2141M}. 
The GW signals are upgraded from 2D to 3D at post-bounce times of 5\,ms for SFHx\_rr, 3\,ms for SFHo\_rr, 31\,ms for SFHx\_nr and 13\,ms for SFHo\_nr.
Graviational redshift is included in the neutrino transport.
 
As discussed above, The SFHo and SFHx EoS \citep{2013ApJ...774...17S} are used at high density.
At low densities, we use an EoS accounting for photons, electrons, positrons and an ideal gas of nuclei together with a flashing treatment for nuclear reactions \citep{2002A&A...396..361R}. 
 
The models have a grid resolution of $550\times 128 \times 256$  zones in radius, latitude and longitude. The grid reaches out to $10^5$\,km, and 21 energy zones are used in the transport solver. To save computer time, we follow the collapse phase in axisymmetry (2D) and map to 3D shortly after bounce, imposing small random seed perturbation to trigger the growth of non-axisymmetric modes.

\section{Hydrodynamic Evolution}
\label{sec:dynamics} 

\begin{figure}
\includegraphics[width=\columnwidth]{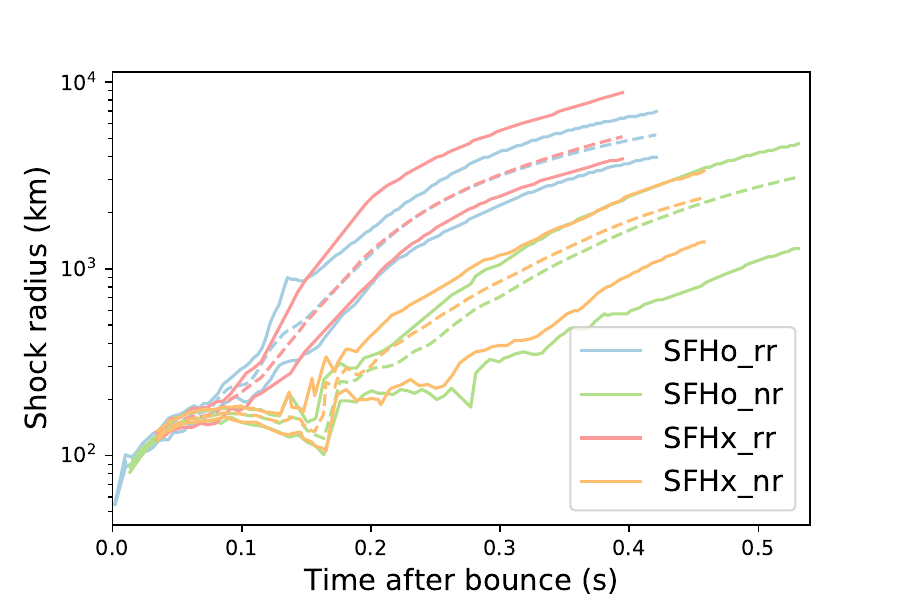}
\includegraphics[width=\columnwidth]{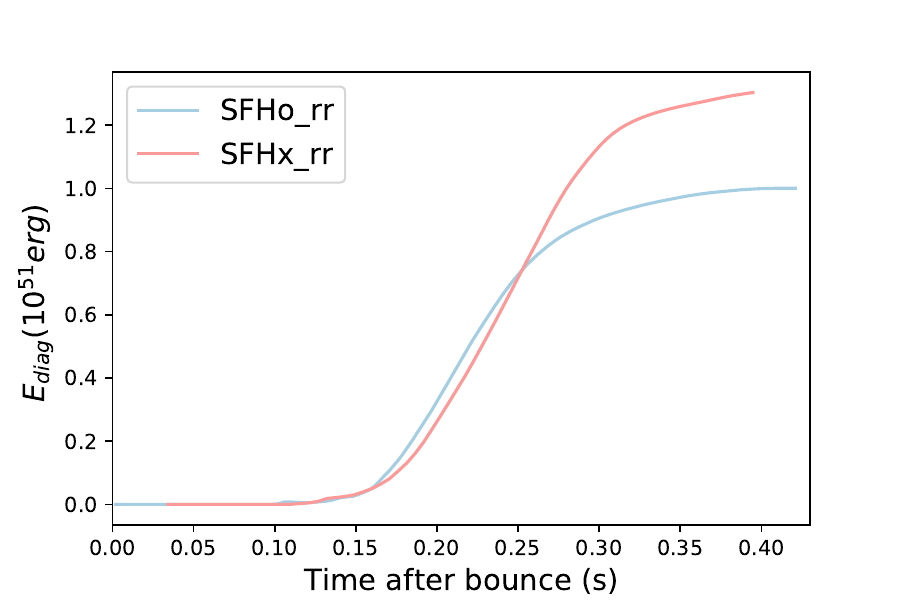}
\includegraphics[width=\columnwidth]{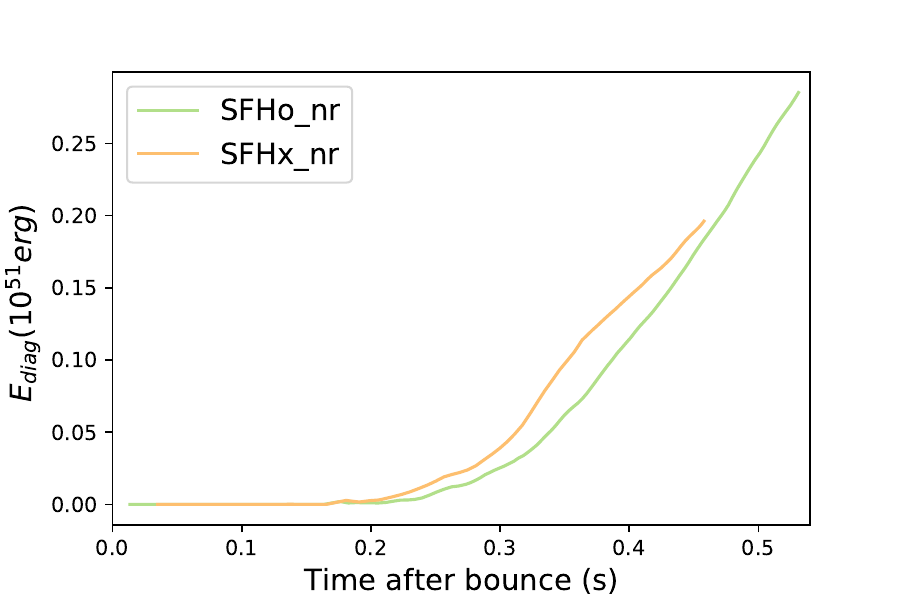}
\caption{(Top) The minimum, maximum, and average shock radius for all models. The shock is revived earlier in the rotating models. (Middle) The explosion energy for the rapidly-rotating models. (Bottom) The explosion energy for the non-rotating models. Rotating models explode earlier with higher explosion energies. }
\label{fig:expl_properties}
\end{figure}

Shock radii for all models are shown in Figure~\ref{fig:expl_properties}. The non-rotating models undergo shock revival at $\sim200$\,ms after bounce, which is reasonably fast but still in the expected range for neutrino-driven explosions.
The time of explosion coincides with the infall of the silicon-oxygen shell interface for this progenitor, which generally improves the neutrino heating conditions, though this was usually not enough to trigger explosions in non-magnetohydrodynamic simulations using this progenitor \citep{mueller_12}. 
Outcomes for the $15M_\odot$ progenitor of
\citet{1995ApJS..101..181W} and other $15M_\odot$
progenitors have been mixed in terms of whether
and when shock revival occurs
\citep{2009A&A...496..475M, 2015ApJ...807L..31L, 2017arXiv170107325Y, andresen_19, 2020PhRvD.102b3027M}. 
The early onset of the explosion in our non-rotating models is not unexpected because of the supportive effect of dynamo-generated magnetic fields \citep{2020MNRAS.498L.109M}.

\begin{figure*}
\includegraphics[width=\columnwidth]{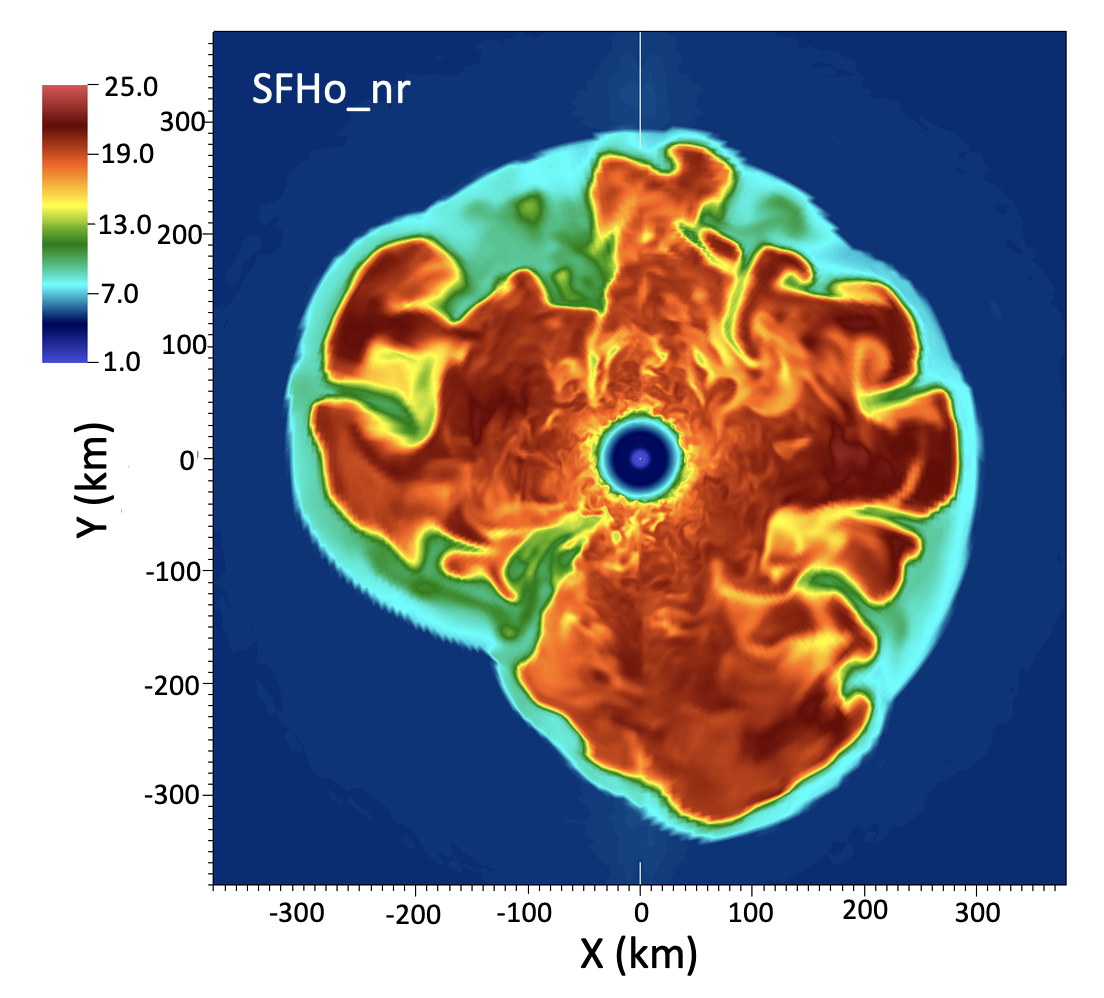}
\includegraphics[width=\columnwidth]{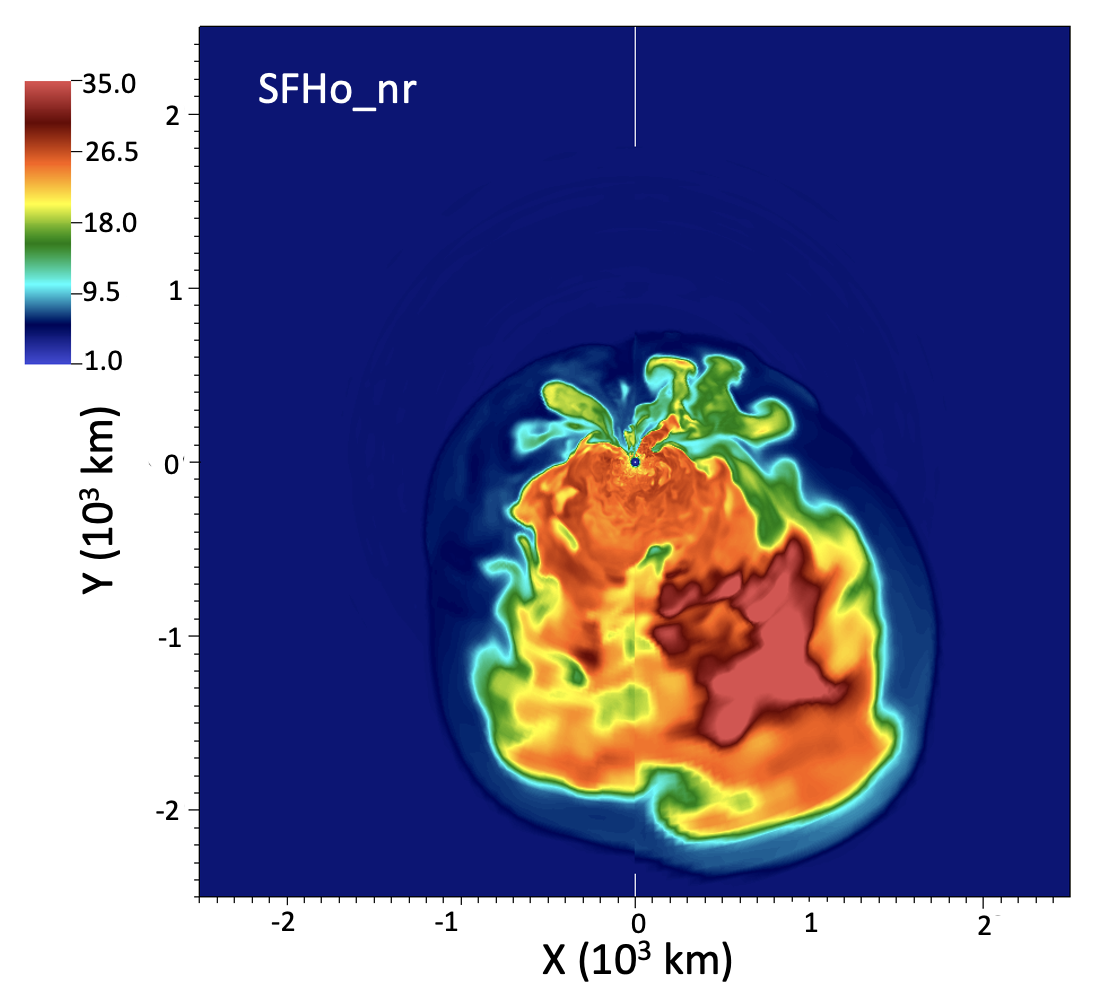}
\includegraphics[width=\columnwidth]{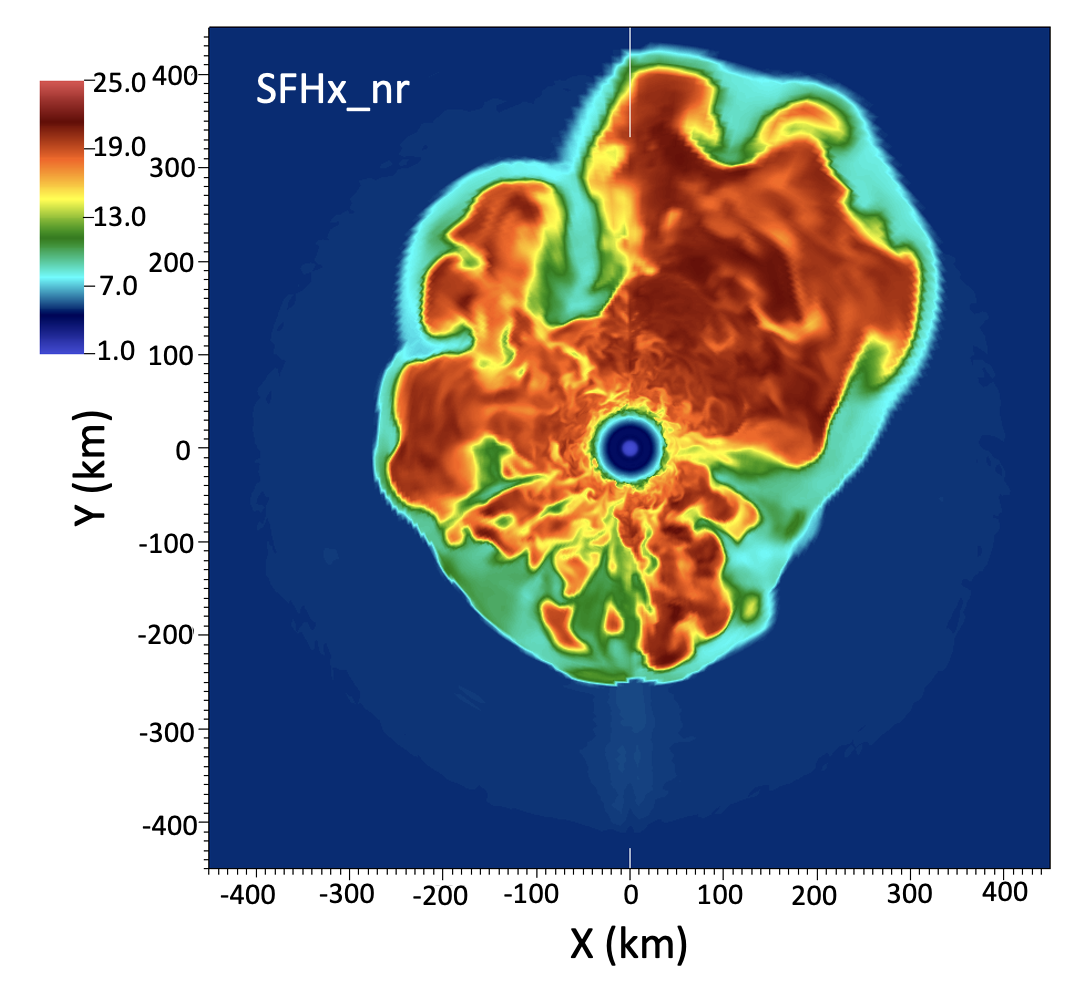}
\includegraphics[width=\columnwidth]{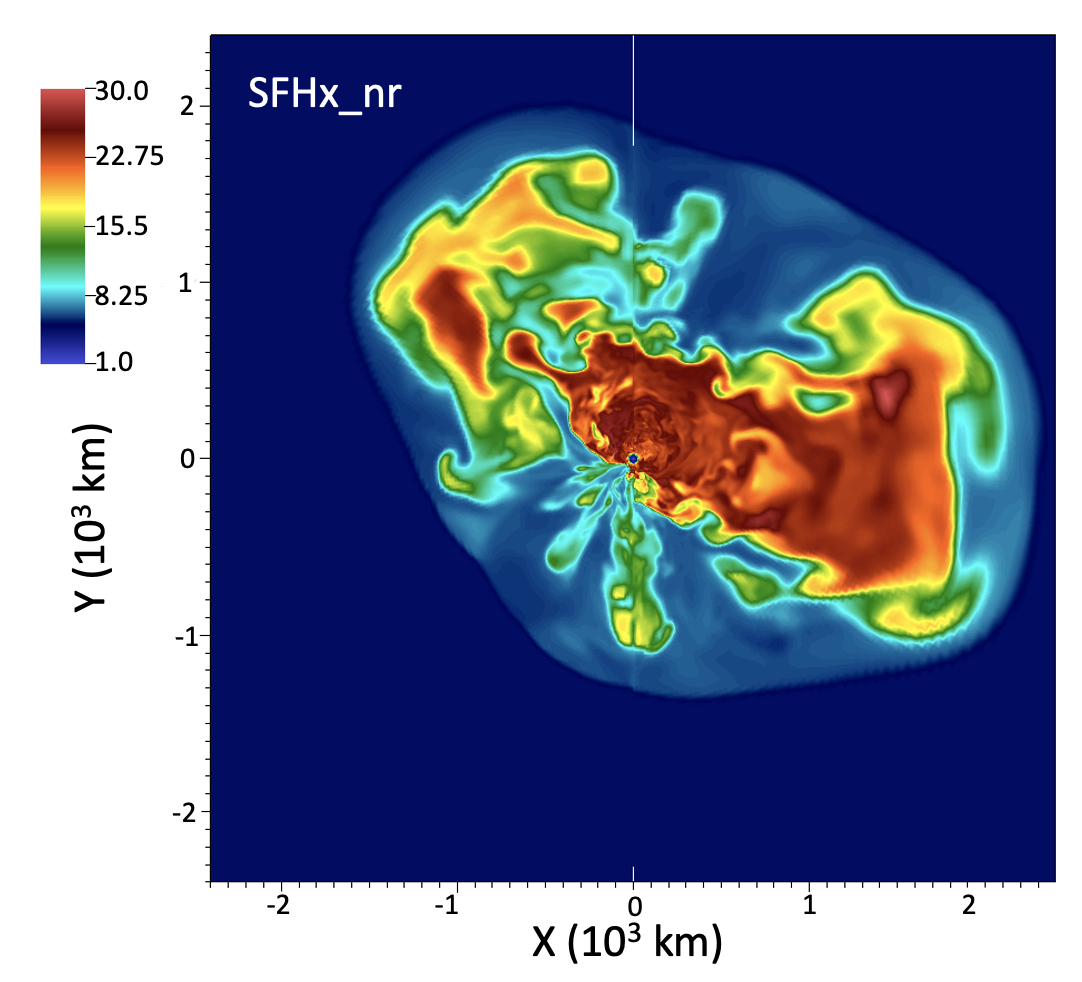}
\caption{  The entropy on 2D meridional slices for the non-rotating models at 200\,ms after bounce (left) and 400\,ms after bounce (right). The top panels show model SFHo\_nr, and the bottom panels show model SFHx\_nr. 
Model SFHo\_nr develops a rather unipolar explosion, whereas model SFHx\_nr falls between a unipolar and bipolar explosion geometry.}
\label{fig:entropy_norot}
\end{figure*}

Figure \ref{fig:entropy_norot} visualises the geometry of the developing neutrino-driven explosions using meridional slices through the simulations.
Around the onset of the explosion, the shock geometry is mildly aspherical in both non-rotating models. Further into the explosion phase, model SFHo\_nr develops a strongly unipolar explosion geometry, whereas SFHo\_nr becomes only a moderately asymmetric explosion that looks neither clearly unipolar or bipolar.

\begin{figure*}
\includegraphics[width=\columnwidth]{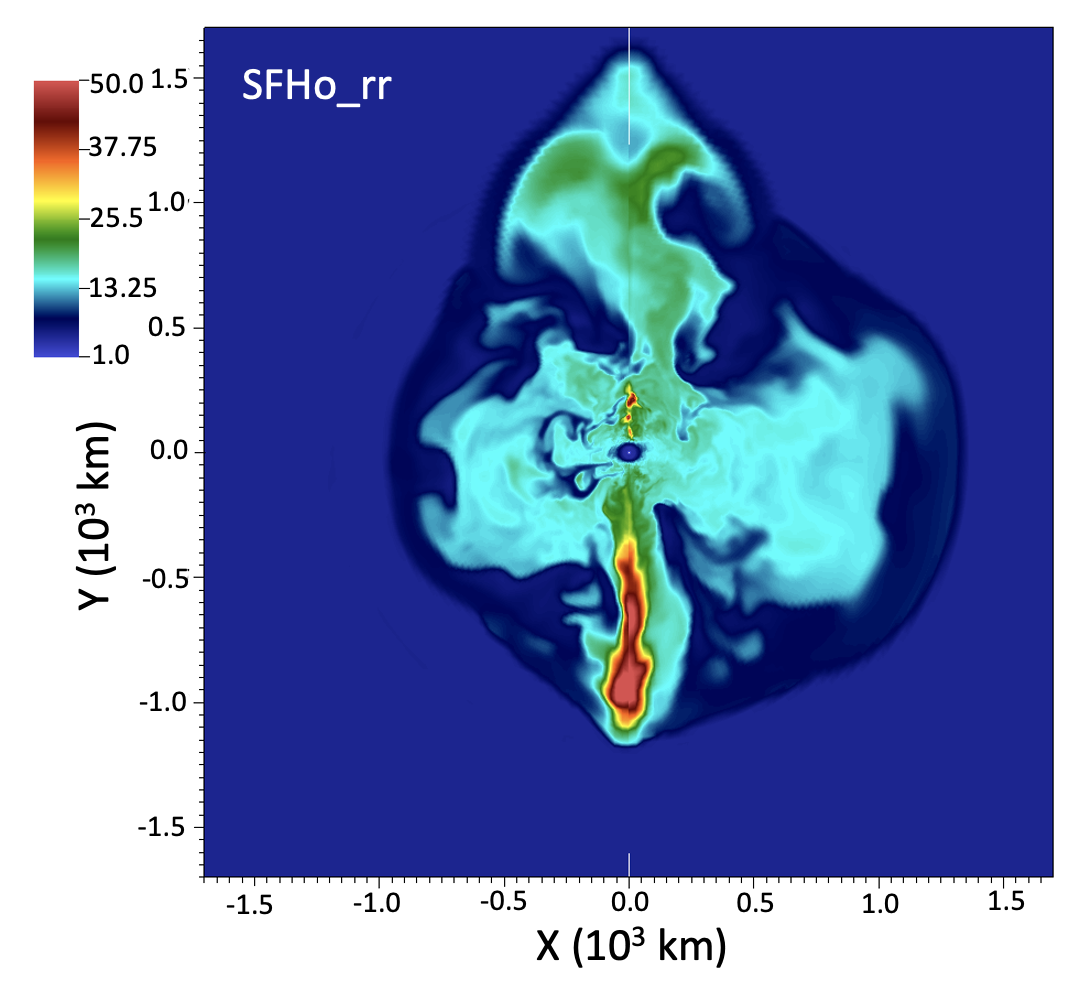}
\includegraphics[width=\columnwidth]{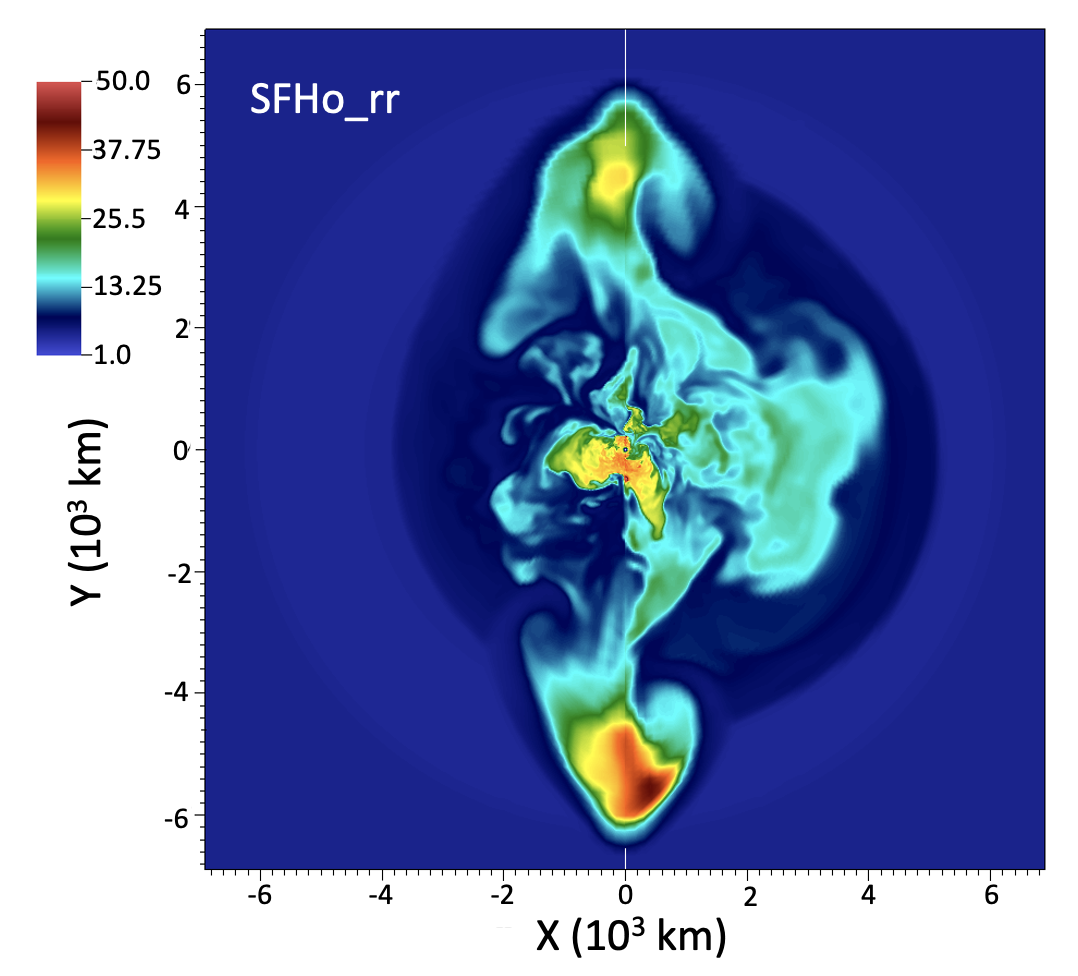}
\includegraphics[width=\columnwidth]{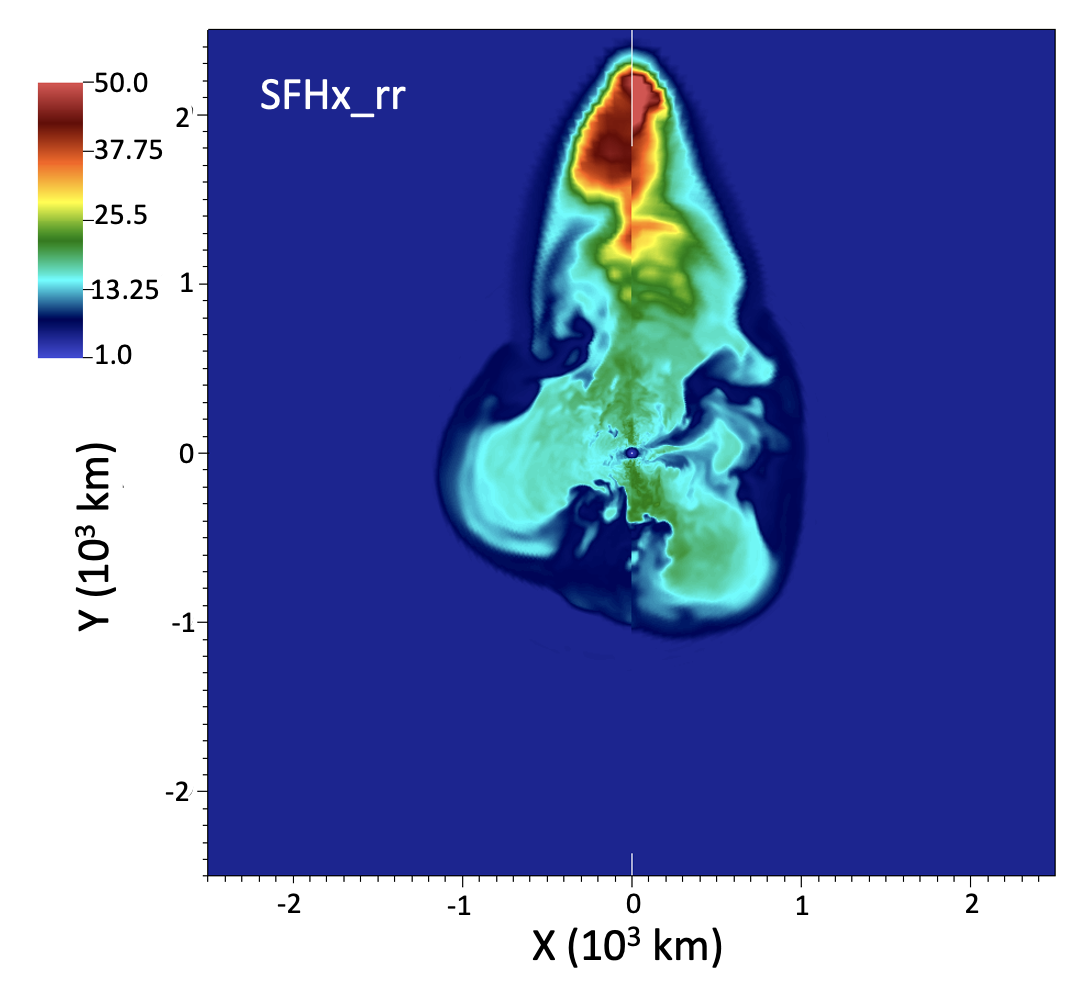}
\includegraphics[width=\columnwidth]{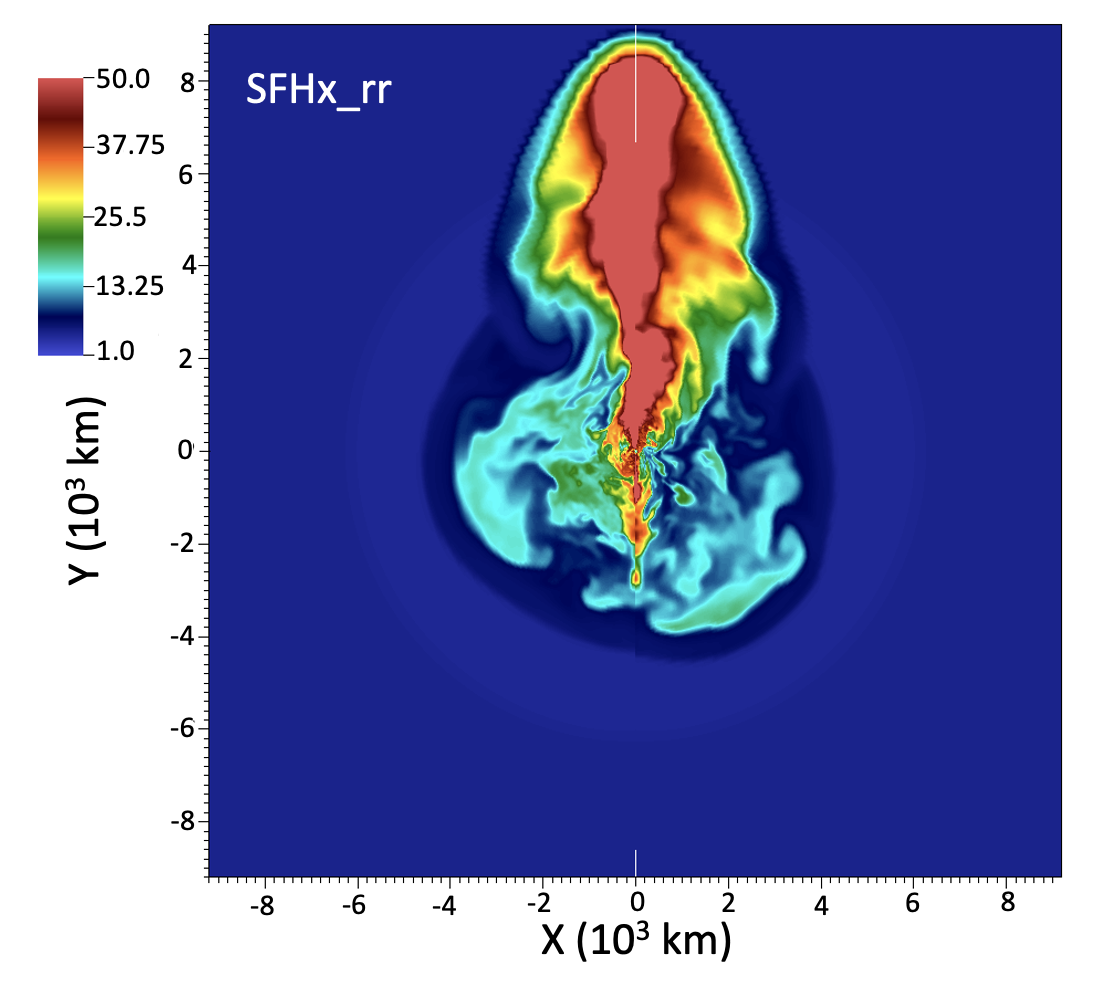}
\caption{ The entropy on 2D meridional slices for the rotating models at 200\,ms after bounce (left) and 400\,ms after bounce (right). The top panels show model SFHo\_rr, and the bottom panels show model SFHx\_rr. The SFHo\_rr model has a clearly visible jets in both directions, but the jets are not stable and strongly distorted or even disrupted by the kink instability. The SFHx\_rr model only has a jet-like outflow in the north direction. }
\label{fig:entropy_rapid}
\end{figure*}

The rapidly rotating models undergo shock revival earlier at $\sim100$\,ms after core bounce. Earlier shock revival times in magnetohydrodynamic simulations of rapidly rotating progenitors is consistent with previous work \citep{2012ApJ...750L..22W, 2014MNRAS.445.3169O, moesta_14b, 2023MNRAS.522.6070P, 2022MNRAS.512.2489O}.  
It is remarkable that such a fast explosion develops even for the low initial field strength assumed in models SFH\_rr and SFHx\_rr. However, a closer look at the explosions reveals that although jet-like outflows are present they are not dominated by strong jets. The evolution of the rapidly rotating models in meridional 2D slices of the entropy, at 200\,ms and 400\,ms post bounce, are shown in Figure~\ref{fig:entropy_rapid}.
The SFHo\_rr model has jet-like structures in both directions, however the jet is larger in the south direction. These jets do not remain collimated, but are very unstable, either because of the kink instability \citep{moesta_14b,bugli_21, bugli_22, 2020arXiv200807205O, 2020MNRAS.492.4613O,2023MNRAS.522.6070P} or because they are buffeted around by other outflows and downflows.
The SFHx\_rr model has no jet in the south direction, and forms a big jet-like structure in the North direction,
which proves robust at later times, but is not strongly collimated and strongly distorted by instabilities.
As the focus of this paper is on the GW emission, we leave a more extensive study of the jet phenomenology for future work. 

There are no significant differences in the shock radius for the different EoS, which is different to our previous neutrino-driven explosions in \citet{2021MNRAS.503.2108P}, where we did observe significantly different shock revival times for the different EoS. Different from the models in \citet{2021MNRAS.503.2108P}, which considered the collapse for very massive cores that quickly results in neutron stars close to the maximum mass of the respective EoS, the current simulations produce neutron stars of rather low mass. For these low masses, the mass-radius relation of the SFHo and SFHx EoS do not differ strongly, so less of an effect on the explosion dynamics is expected.

The diagnostic explosion energies are shown in Figure~\ref{fig:expl_properties},
and the final energies at the end of the simulations are given in Table~\ref{tab:models}. For the non-rotating models, the energy is still growing rapidly at the time that the simulations were ended, which is typical for models with only $\sim 0.5$\,s in duration, and they reach a final value similar to what is found in other simulations of non-rotating neutrino-driven models with a similar progenitor mass. For the rotating models, the energy grows very rapidly after shock revival but then quickly levels off after a few hundred milliseconds, as seen in previous models with stronger magnetic fields \citep{varma_22, 2023MNRAS.522.6070P}.
The final energy is significantly smaller than our previous magnetorotational explosions
in \citet{2023MNRAS.522.6070P}, which had higher progenitor masses and stronger initial magnetic fields. The explosion energies are similar to what is typically observed in type Ib or Ic supernovae, but not in hypernovae. 
While purely neutrino-driven explosion models are able
to reach explosion energies of $10^{51}$\,erg, they can currently do this only on significantly longer time scales
of seconds instead of hundreds of milliseconds \citep{mueller_17,2021ApJ...915...28B, 2023ApJ...957...68B}. 
Our rapidly rotating models can be seen as exploring the potential GW emission from supernovae with typical energies, but faster powering of the explosion than predicted by current neutrino-driven simulations.

We also investigate the impact of the choice of the EoS on the shock revival time and energy. The SFHx\_rr model has significantly larger final explosion energy than model SFHo\_rr. This is different to our previous results of neutrino-driven explosions using the same two EoS. In \citet{2021MNRAS.503.2108P}, we found that the model with the SFHx EoS underwent shock revival $\sim100$\,ms later than the SFHo model, and reached a significantly lower final explosion energy.  The differences in explosion energies are likely due to stochastic variations as we do not see any evidence for progenitor dependence.

\begin{figure*}
\includegraphics[width=0.95\columnwidth]{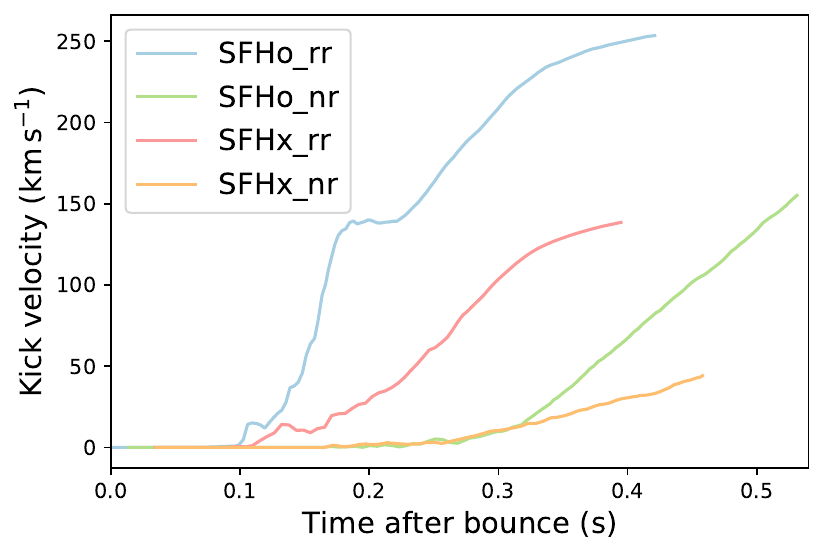}
\includegraphics[width=0.95\columnwidth]{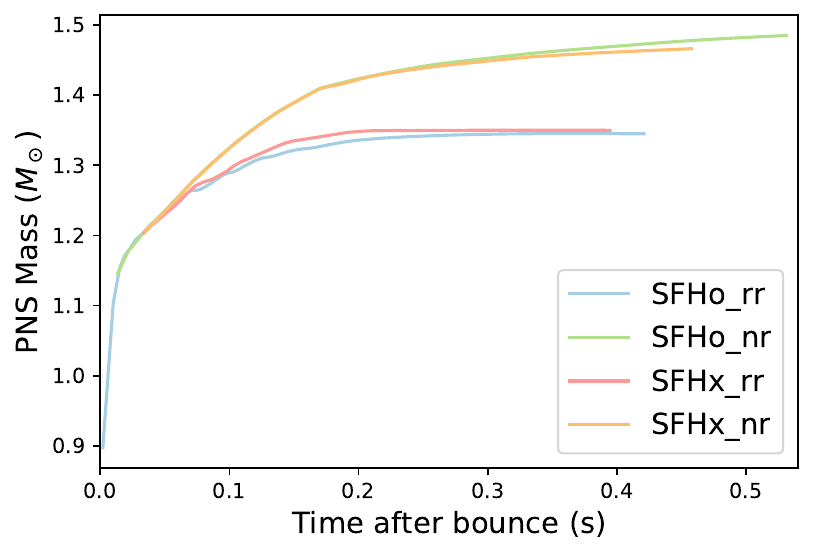}
\includegraphics[width=0.95\columnwidth]{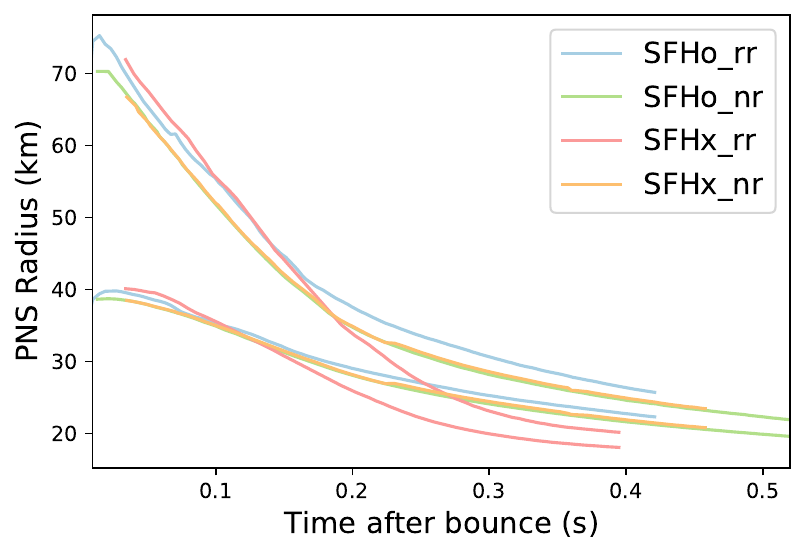}
\includegraphics[width=\columnwidth]{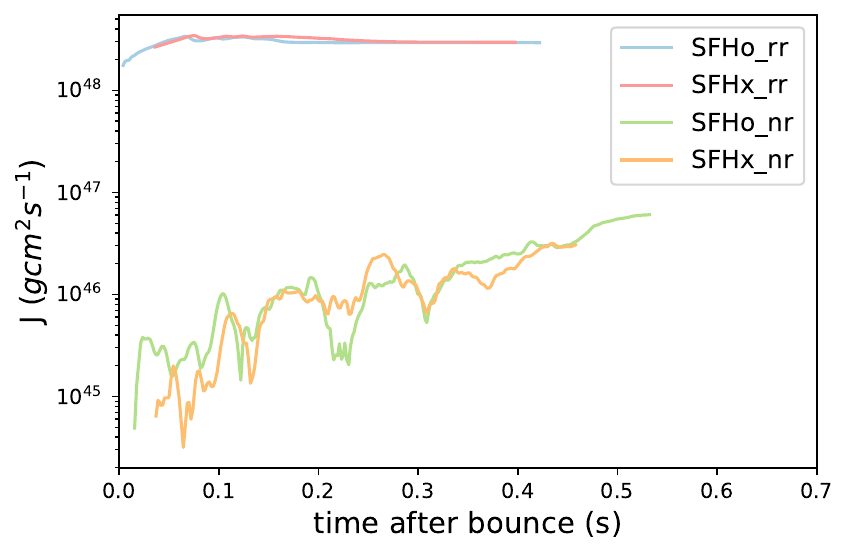}
\caption{The top left panel shows the kick velocities for all of the models.
Models with the SFHo EoS achieved higher kick velocities than models with the SFHx EoS. The kick is still growing by the end of the simulation time, especially
for models SFHo\_nr and SFHx\_nr.
The top right panel shows the baryonic PNS masses for all the models. Rapid progenitor rotation results in a lower final mass of the PNS. The bottom left shows the proto-neutron star radii. SFHx\_rr has a smaller radius than the other models. The bottom right panel shows the PNS angular momentum. }
\label{fig:pns_properties}
\end{figure*}

Although our simulations finish too early to obtain accurate final measurements of the PNS properties at least for the non-rotating models, we show the PNS masses, radius, spin and kicks  in Figure~\ref{fig:pns_properties}.
Kicks develop faster for the rapidly rotating models due to the earlier onset of the explosion, but they asymptote more quickly at moderate values of $254\,\mathrm{kms}^{-1}$ and $139\,\mathrm{kms}^{-1}$ for the SFHo\_rr and  SFHx\_rr models respectively. Final kicks for the non-rotating models cannot be determined yet, but at least for SFHo\_nr, the kick may well end up higher than for the rapidly rotating models. Variations in kick magnitude are likely determined by the stochastic variations of the explosion geometry evident from Figure~\ref{fig:entropy_rapid} and \ref{fig:entropy_norot}.

All of the models asymptote quickly towards a final baryonic mass value. Rapid progenitor rotation results in a lower final PNS mass due to earlier shock revival, which is consistent with previous work, and there is not a significant difference in final PNS mass values between the different EoS. However, the PNS radius shrinks significantly faster for model SFHx\_rr than for model SFHo\_rr. 
The fast shrinking is ultimately caused by a slightly different angular momentum distribution
in the PNS. In model SFHx\_rr, the positive angular momentum gradient in the outer shells of
the PNS is strong enough to inhibit convection. This reduces the PNS convection zone to about
half the width found in model SFHo\_rr. 
It is not easily possible to establish the causal factors behind this difference because of feedback processes that may result in a bifurcation. EoS properties directly influence stability through the coefficient of the electron fraction gradient in the Ledoux criterion
(see \citealp{jakobus_24}, Equation~21). In addition, EoS-dependent neutrino opacities affect the evolution of the entropy and electron fraction and hence convective stability. If convective instability is sufficient to overcome stabilisation by angular momentum gradients, convective transport of angular momentum can reinforce instability and lock the PNS convection into an active state. If convective instability becomes too weak to overcome the stabilising angular momentum gradient, redistribution of angular momentum will be slower and convection may be unable to restart.
The rapidly rotating models maintain very similar angular momentum after the onset of shock revival until the end of the simulation. There
is some spin-down of the PNS early on around shock revival, but the angular momentum remains high of order $\gtrsim 2.5\times 10^{48}\,\mathrm{erg}$,
and a substantial amount of rotational energy could still be released on longer time scales. This is in contrast to the magnetorotational explosion models of \citet{2023MNRAS.522.6070P}, where little rotational energy is left in the PNS already quite early during the explosion. The non-rotating models gain angular momentum over time due to stochastic accretion.

\section{Gravitational Wave Properties}
\label{sec:grav_waves_time} 

\begin{figure*}
\includegraphics[width=\textwidth]{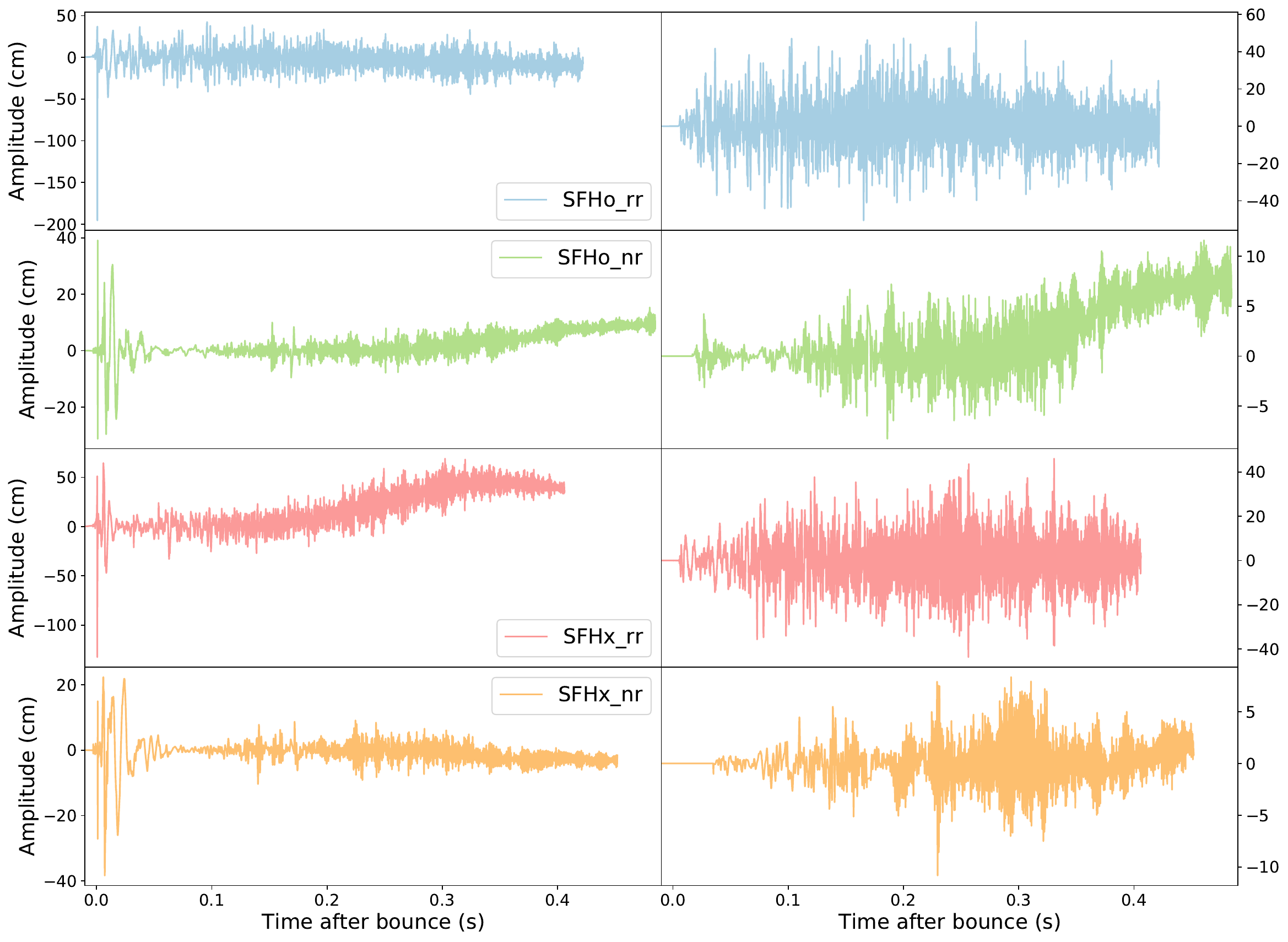}
\caption{The time series of the GW signal from matter for all models as measured for an observer at the equator ($\theta=\phi=90$). The left column shows the plus polarisation, and the right column shows the cross polarisation. The rapidly rotating models have higher GW amplitudes. Some of the models have started to develop low-frequency tails before the end of the simulation time. }
\label{fig:waveforms}
\end{figure*}

The time series of the GWs from matter as measured by an observer in the equatorial plane ($\theta=\phi=90^\circ$), are shown in Figure~\ref{fig:waveforms}. 
As the core-bounce time was simulated in 2D, the non-rotating models are showing strong ringdown after prompt convection at the start, and slowly evolve to lower GW amplitudes as the initial axisymmetry is lost in favour of less coherent motions in 3D. This has the largest impact on model SFHo\_nr, which has a high amplitude of 10\,cm at the start in the plus polarisation. This issue does not have an impact on the rotating models where there is a (physically) preferred axis anyway. 

The non-rotating models reach GW amplitudes of $\sim 10$\,cm. This is similar to typical amplitudes from our previous neutrino-driven explosion models, for example, model s18 from \citet{2019MNRAS.487.1178P}, and model y20 from \citet{2020MNRAS.494.4665P}. This shows that the addition of weak magnetic fields does not have a big impact on GW amplitudes. The rotating models reached GW amplitudes of $\sim 40$\,cm. This amplitude is similar to our previous rapidly rotating model m39 \citep{2020MNRAS.494.4665P}, which had no magnetic fields, but significantly smaller than our previous models with strong magnetic fields and rapid rotation, models m39\_B10 and m39\_B12 from \citet{2023MNRAS.522.6070P}. This is consistent
with the much lower explosion energy in the current models compared to $\sim 3\times 10^{51}\,\mathrm{erg}$ in m39\_B10 and m39\_B12,
and
suggests that a stronger magnetic field strength is required for powerful GW signals from magnetorotational explosions. 

In the time series, small low-frequency tails from matter motions are observed in the SFHx\_rr and SFHo\_nr models due to asymmetric shock expansion
\citep{2009ApJ...707.1173M,2013ApJ...766...43M}, and also in model SFHx\_nr after $\sim0.4$\,s. Previous models have only started to deviate significantly from zero after $\sim0.5$\,s \citep{2019MNRAS.487.1178P, 2020MNRAS.494.4665P}, so longer simulations of these models would be required for a good estimation of the final amplitude of the low-frequency emission due to matter motions. 

\begin{figure*}
\includegraphics[width=\columnwidth]{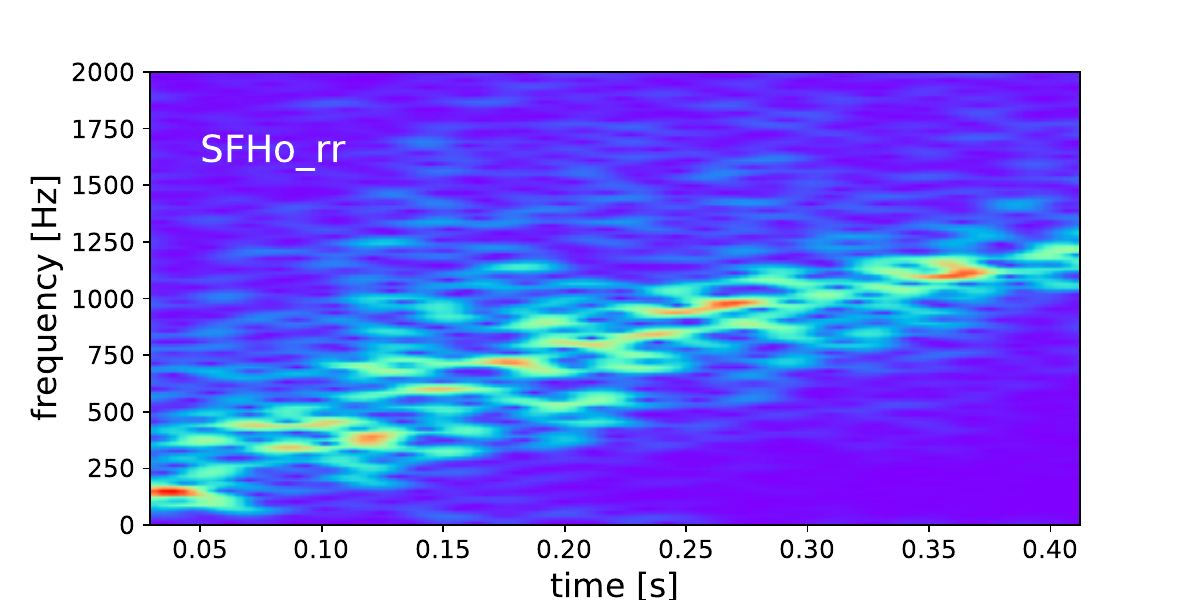}
\includegraphics[width=\columnwidth]{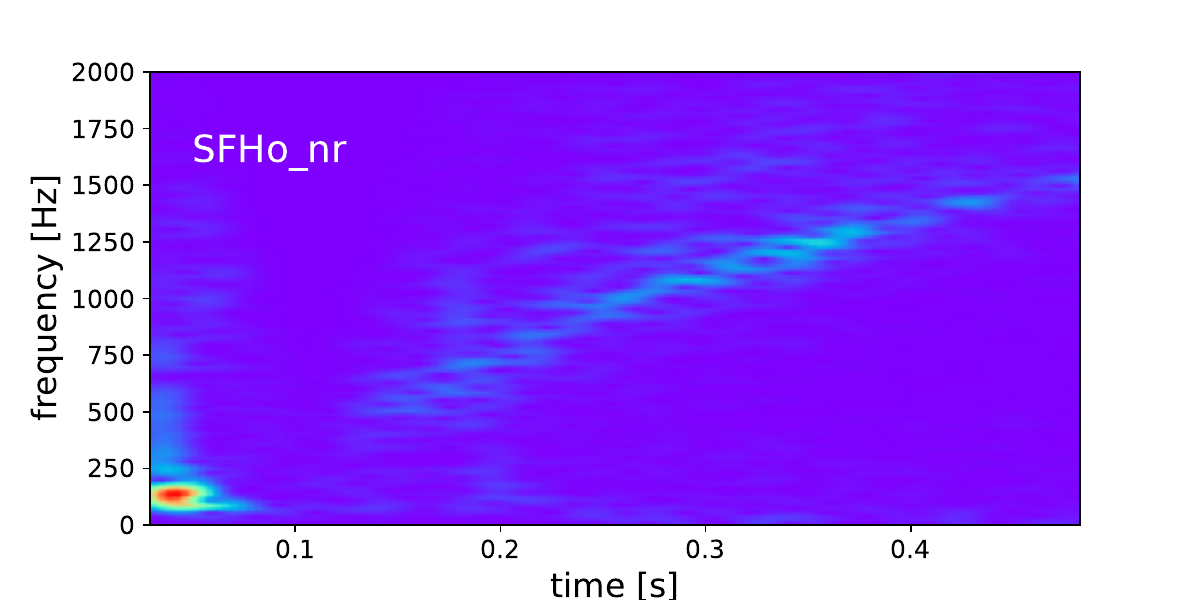}
\includegraphics[width=\columnwidth]{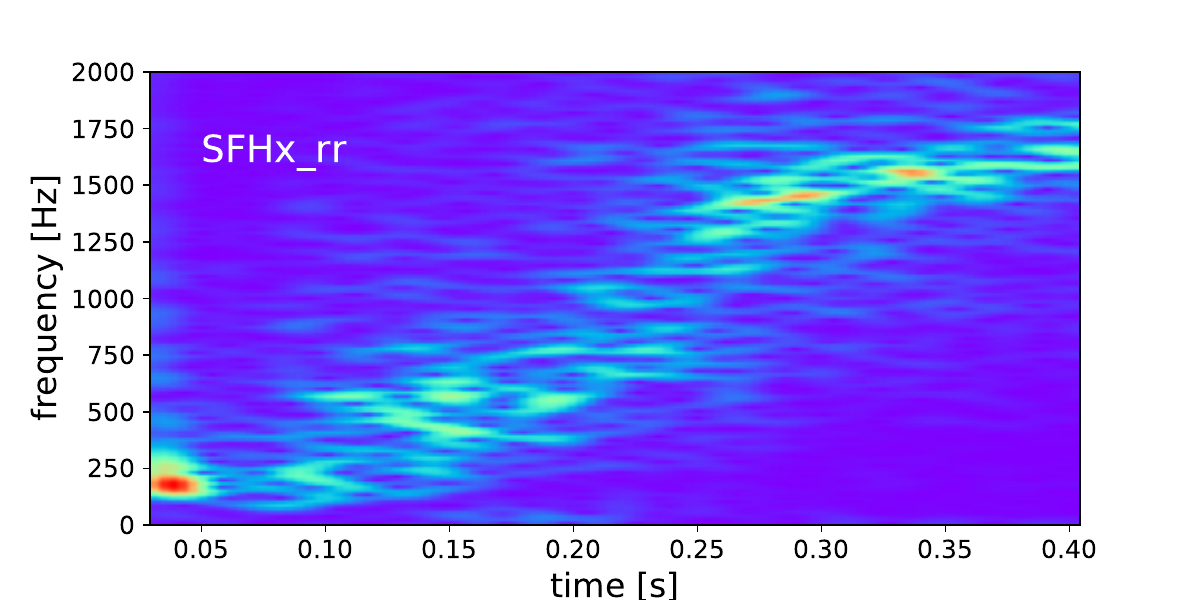}
\includegraphics[width=\columnwidth]{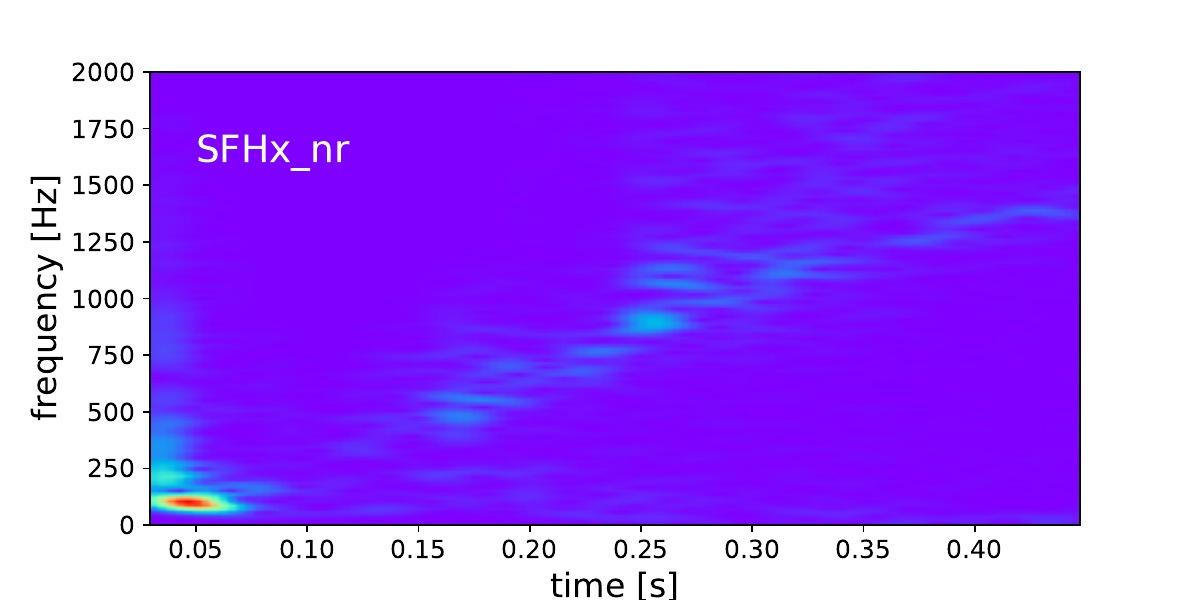}
\caption{Spectrograms of 
$h_+ + h_\times$
for all our GW signals. The top left panel shows model SFHo\_rr, the top right panel shows model SFHo\_nr, the bottom left panel shows model SFHx\_rr, and the bottom right panel shows model SFHx\_nr. The non-rotating models show some low frequency GW emission from the SASI before  shock revival time. All models reach high frequencies of above 1000\,Hz before the end of the simulation. Model SFHx\_rr reaches higher frequencies due to its smaller PNS radius. }
\label{fig:spectrograms}
\end{figure*}

Spectrograms of the GW emission are given in Figure~\ref{fig:spectrograms}. At early times, the  
non-rotating models have a high-power hotspot at low frequency caused by the ringdown after the transition from 2D to 3D. In the non-rotating models, some low-frequency emission due to the SASI is visible before 0.2\,s when shock revival stops the SASI activity. The non-rotating models both have a similar high-frequency f/g-mode emission. They both reach a maximum frequency of $\sim 1500$\,Hz by the end of the simulation. The GW frequency is higher than in our previous simulations because of the pseudo-Newtonian gravity used in our simulation. 
Previous studies have shown that the GW frequency is about $\sim 20$\% higher than when full general relativity is used \citep{2013ApJ...766...43M}, due to missing
relativistic correction terms in the Brunt-V\"ais\"al\"a frequency (including, but not limited to a time dilation factor from the lapse function $\alpha$, see Equations~16 and
C8 in \citealp{2013ApJ...766...43M}) in the pseudo-Newtonian approximation.
The rapidly rotating models have no visible lower frequency modes. Their high-frequency emission has a different shape and maximum frequency for the two different EoS, with a slight S-shape for model SFHx\_rr. This is due to the difference in the PNS radius between the two models, as shown previously in Figure \ref{fig:expl_properties}, where the smaller final PNS radius for model SFHx\_rr has resulted in a higher maximum frequency of the f/g-mode mode.  The dominant emission band is distinctly broader for the models with rapid rotation, with a width of $\sim 500$\,Hz instead of $\sim 250$\,Hz, reminiscent of the cases with rapid rotation and strong fields in the 2D study of
\citep{Jardine_2021}.
Such a broadening of the dominant emission band may be an indicator for rapid rotation in the event of a GW detection from a Galactic CCSN.

\section{The Neutrino Memory}
\label{sec:neutrino} 

\begin{figure*}
\includegraphics[width=\textwidth]{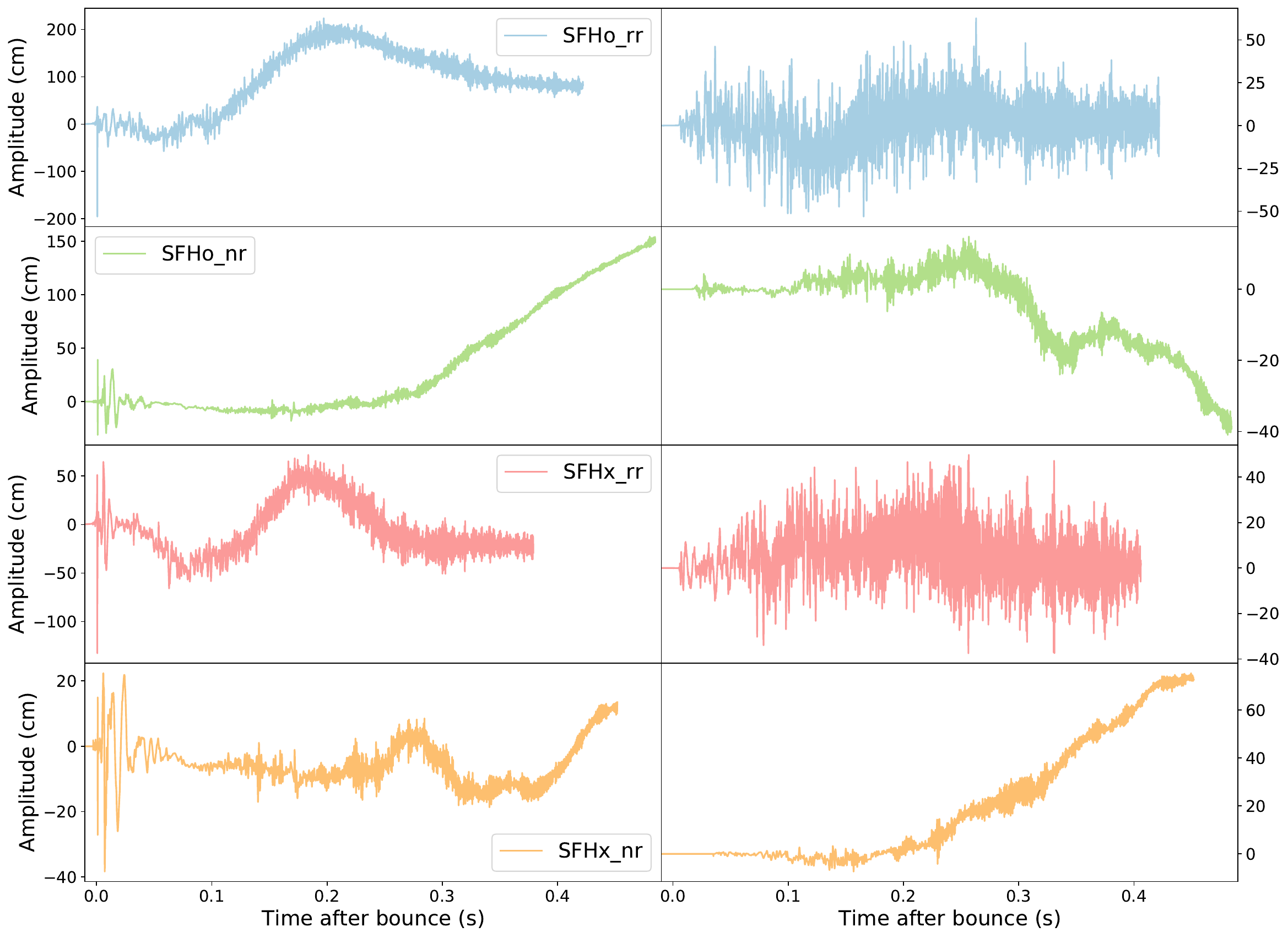}
\caption{The GW amplitude time series from both matter and anisotropic neutrino emission. The left column shows the plus polarisation, and the right column shows the cross polarisation, both calculated at $\theta=\phi=90^\circ$ (equatorial observer). All models show a significant increase in amplitude at low frequencies (slow variations) compared to the matter signal due to the neutrino emission anisotropy. }
\label{fig:neu_matter}
\end{figure*}

In addition to the low-frequency component to the GW emission from matter, there generally is a high-amplitude low-frequency component to the GW emission from the asymmetric emission of neutrinos (neutrino memory, \citealp{1978ApJ...223.1037E}).
The neutrino memory will usually dominate over the low-frequency matter signal 
\citep{marek_09,2013ApJ...766...43M}.
This component of the GW signal is often ignored in CCSN simulations that focus on GWs, as it is outside of the frequency band for the current GW detectors. However, understanding the GW signals at frequencies below 10\,Hz is important for the science case for the next generation of GW detectors. In this section, we show how the GW emission changes when the signal from the asymmetric emission of neutrinos is also included. 

To calculate the GW signal from anisotropic emission of neutrinos, we use Equation~(24) from \citet{1997A&A...317..140M},
which gives the transverse-traceless perturbations
$h_{ij}^\mathrm{TT}$ of the metric from neutrino memory,
\begin{equation}
h_{ij}^\mathrm{TT} (\mathbf{X},t) = \frac{ 4G }{ c^4R} \int_{-\infty}^{t-R/c} dt' \int_{4\pi}
d\Omega' \frac{ (n_in_j)^\mathrm{TT} }{ 1-\cos\theta } \cdot 
\frac{d L_{\nu}(\mathbf{\Omega}',t') }{d\Omega'} ,
\end{equation}
where $G$ and $c$ are gravitational constant and the speed of light, $R$ is the distance to the source, $\theta$ is the angle between the direction towards the observer and the direction $\Omega'$ of the radiation emission, $d L_{\nu}(\mathbf{\Omega}',t') / d\Omega'$ is the direction-dependent neutrino luminosity, which is the energy radiated at time $t$ per unit of time and per unit of solid angle into direction $\Omega$. 

The combined GW emission from both matter and neutrinos is shown in Figure~\ref{fig:neu_matter}. 
The low-frequency GW signal from asymmetric neutrino emission has significantly higher amplitude than the matter signal alone. All models show a significant increase 
in GW amplitude when the neutrino memory is included. Model SFHo\_rr reaches amplitudes of 200\,cm in the plus polarization, model SFHo\_nr reaches 140\,cm, model SFHx\_rr reaches 100\,cm, and model SFHo\_nr reaches -150\,cm. Due to the short simulation time of these models, we can only predict the low-frequency emission down to about 2\,Hz, as the minimum measurable frequency is determined by the simulation duration. The fact that the models quickly reach high amplitudes at low frequencies reflects that the models develop large-scale asymmetry in the explosion ejecta soon after the shock is revived. There is slightly less anisotropy in model SFHx\_rr than in the others. The GW signal is still growing significantly in time for the non-rotating models at the end of the simulation time. Longer duration simulations by other groups have shown that the neutrino memory can grow to over a 1000\,cm at later times \citep{2023PhRvD.107j3015V}. This shows the need to carry out long duration simulations to fully capture the both the minimum frequency and maximum amplitude of low-frequency GW emission. 

\begin{figure*}
\includegraphics[width=\columnwidth]{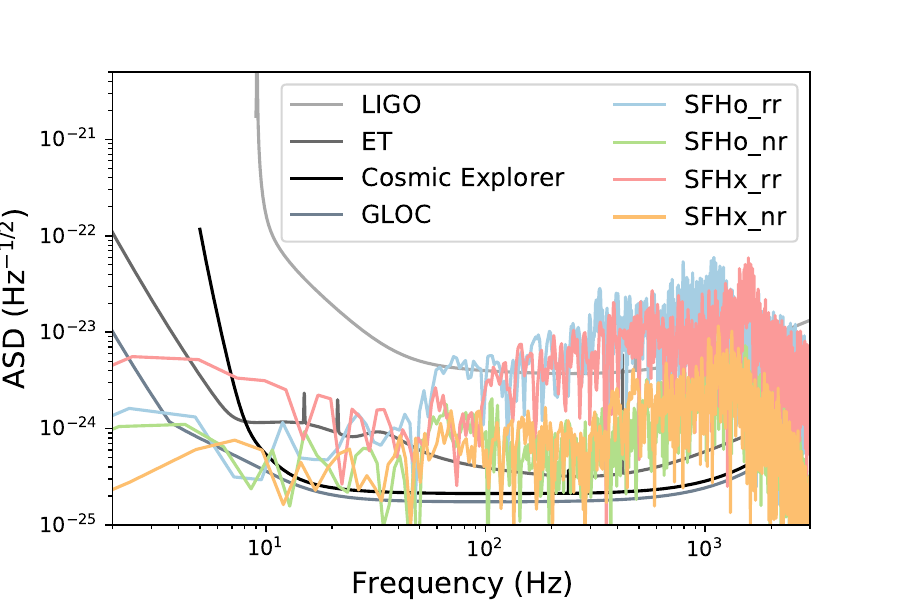}
\includegraphics[width=\columnwidth]{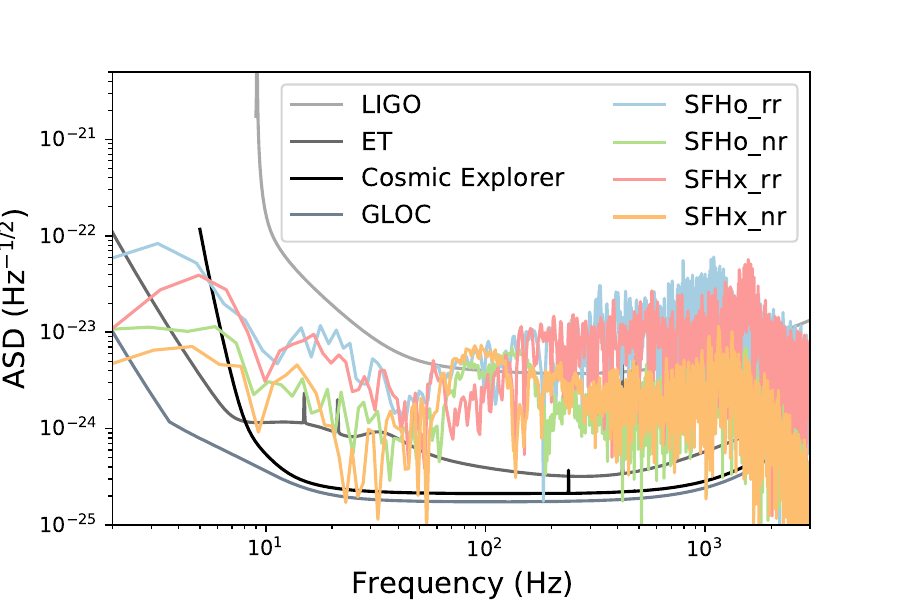}
\caption{ The amplitude spectral density (ASD) of all models, as measured at a distance of 100\,kpc, and the sensitivity curves of the LIGO, Einstein Telescope (ET), Cosmic Explorer and moon-based (GLOC) GW detectors. The left panel shows the matter component only. The right panel shows the spectrum from the combined matter and neutrino signal. The GW spectrum from matter peaks at frequencies of above 1000\,Hz, and the GW spectrum from asymmetric neutrino emission peaks below 10\,Hz. }
\label{fig:gw_asd}
\end{figure*}

We also show the amplitude spectral densities for all models, with the matter components only, and both the matter and neutrino components in Figure~\ref{fig:gw_asd}. 
Including the neutrino memory results
in a much larger amplitude at low frequencies, in the frequency detection bands of Einstein Telescope and proposed moon-based GW detectors. 

\section{Gravitational Wave Directional Dependence and detectability}
\label{sec:direction} 

\begin{figure*}
\includegraphics[width=\columnwidth]{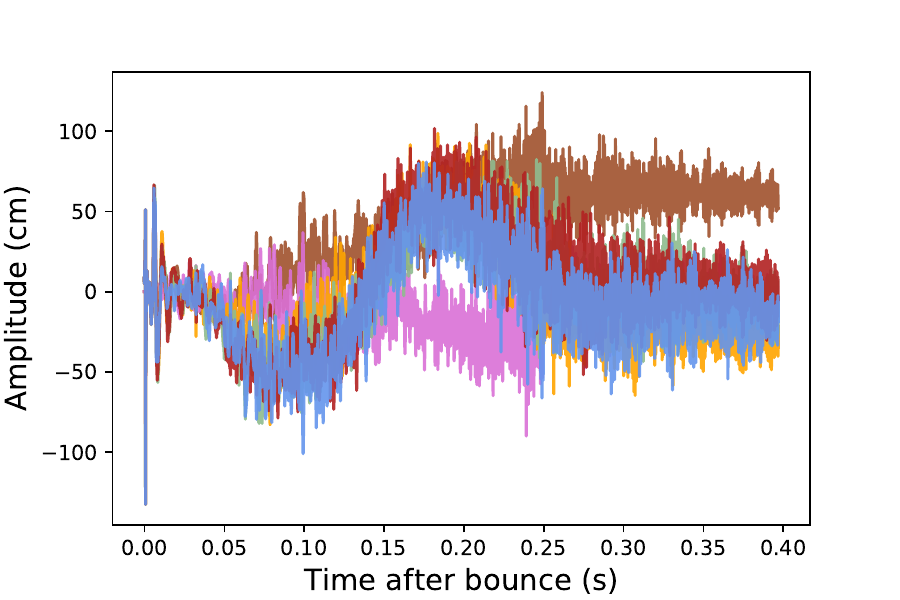}
\includegraphics[width=\columnwidth]{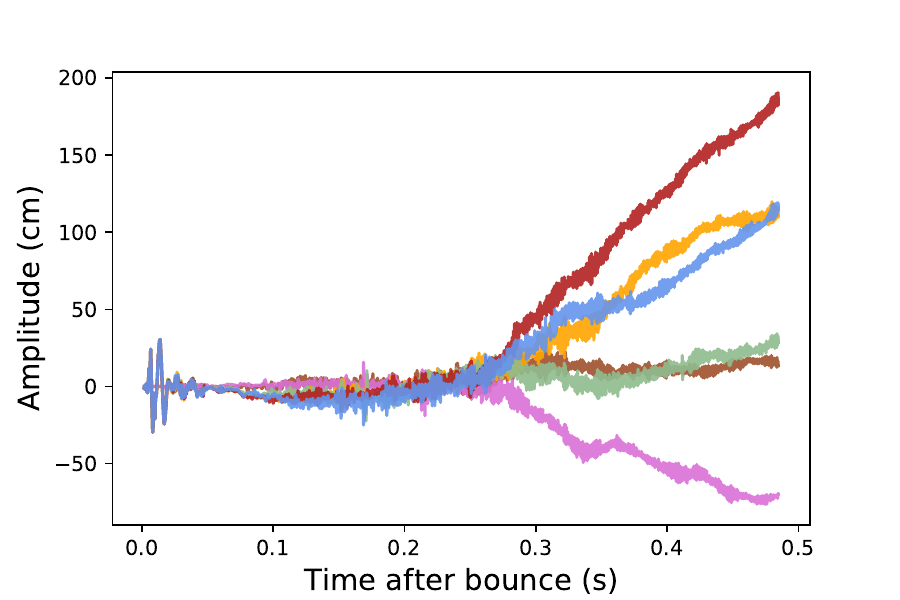}
\includegraphics[width=\columnwidth]{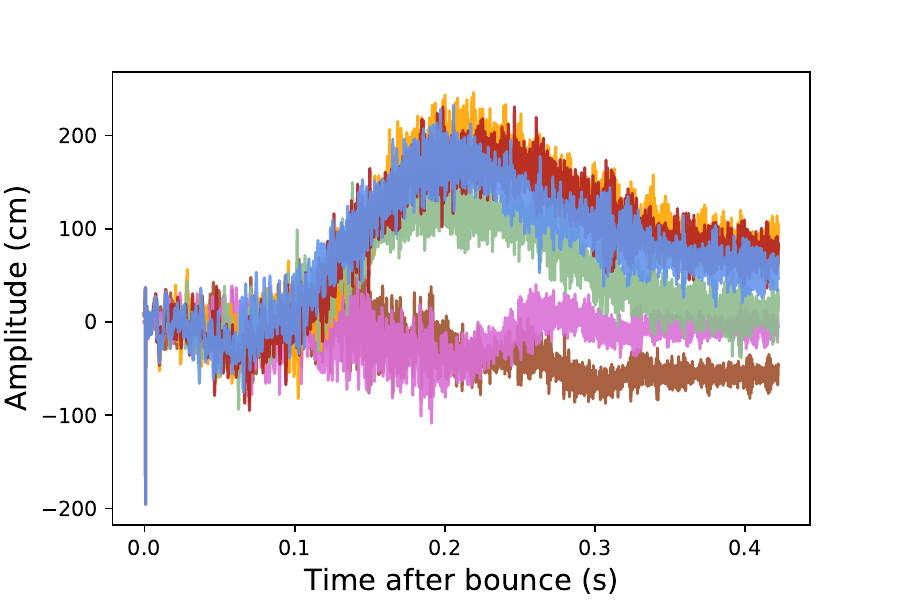}
\includegraphics[width=\columnwidth]{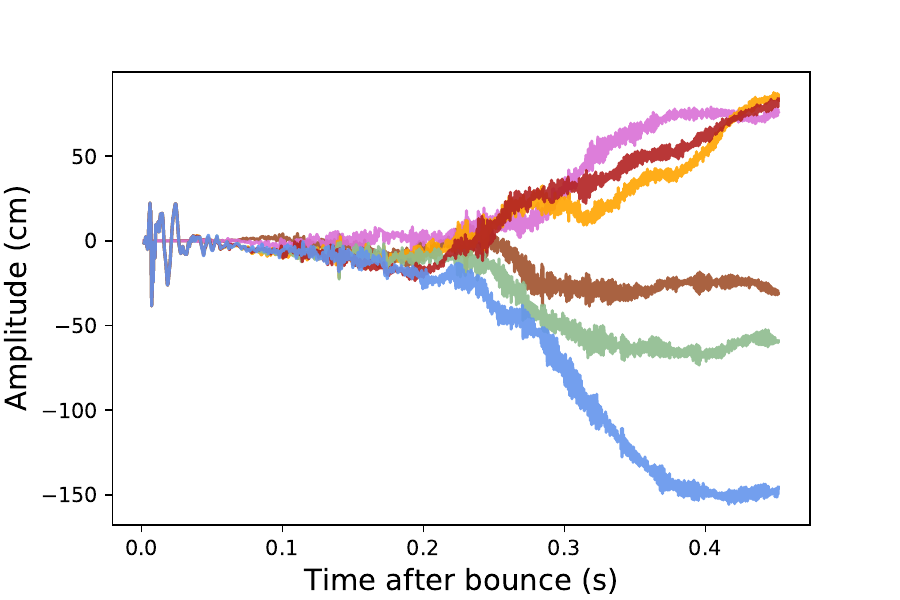}
\caption{ The GW time series of all four models as measured at six different observer angles. (top left) The GWs for model SFHo\_rr. (top right) The GWs for model SFHo\_nr. (bottom left)  The GWs for model SFHx\_rr.  (bottom right) The GWs for model SFHx\_nr. The amplitude of the low-frequency GW emission can vary significantly at different source angles. 
 }
\label{fig:gw_angles}
\end{figure*}

In Figure~\ref{fig:gw_angles}, we show the GW emission calculated 
at six different representative source
angles
$(\theta,\phi)
\in \{
(0^\circ,0^\circ),
(180^\circ,0^\circ),
(90^\circ,0^\circ),
(90^\circ,90^\circ),
(90^\circ,180^\circ),
(90^\circ,270^\circ)\}$.
When including only frequencies in the sensitivity band of current GW detectors, there is no significant difference in the time series of the GW signals at different observer angles. However, when frequencies below 10\,Hz are included, the figure clearly shows a significant difference in amplitude between the different angles.  

To determine if the observer angle has a significant impact on the detectability of the GW signals, we calculate their maximum detection distances at the 6 different angles for a variety of different GW detectors. We define the maximum detectable distance as the distance required for an optimal signal to noise ratio (SNR) of 8. We include the Advanced LIGO detector at design sensitivity, the Cosmic Explorer detector \citep{2021arXiv210909882E}, the Einstein Telescope \citep{2011CQGra..28i4013H}, and the proposed moon-based GW detector, GLOC \citep{2020arXiv200708550J}. The noise curves for the different detectors are shown in Figure~\ref{fig:gw_asd}. For the Einstein Telescope, we use the D configuration \citep{2011CQGra..28i4013H}. For the lunar detector, we use the conservative noise curve from \citet{2020arXiv200708550J}. Due to our short duration simulations, we consider a minimum frequency of 2\,Hz.

The results are shown in Table~\ref{tab:detection}. We show the frequency at which the GW amplitude peaks. This is because the GW detectors sensitivity is frequency-dependent, and the central frequency is also measured by low-latency GW unmodelled searches. For models SFHo\_nr, SFHx\_rr, and SFHx\_nr, 
the GW amplitude peaks shortly after shock revival, and the frequency at which the peak emission occurs varies by a few hundred Hz depending on the orientation angle, which is similar to our previous models. 
However, model SFHo\_rr is different, as the GW signal is fairly uniform in amplitude 
for the duration of the entire signal, resulting in a much larger variance in the peak frequency. This is similar to, for example, some models in \citet{bugli_22}. The frequency along the f/g-mode at which the GW emission peaks can vary a lot between different CCSN simulation codes, as well as within our own simulations. 

For the Advanced LIGO detector, we show the maximum detection distance for the matter component only, as the neutrino component of the signal is of too low frequency to be detectable in Advanced LIGO. The non-rotating models distance varies by about 5\,kpc depending on the source orientation. However, for all orientations, the signals could be detected throughout our Galaxy. The rotating models can be detected out to $\sim 100$\,kpc for SFHx\_rr and 130\,kpc for SFHo\_rr. The detectable distances for the rotating models have a larger difference at different source orientations than the non-rotating models. 

For the Einstein Telescope, the maximum detection distances are above a Mpc for the rotating models, and a few hundred kpc for the non-rotating models. The detection distances increase by as much as a few hundred kpc when the neutrino component of the GW signal is included. This shows that including the GWs due to the asymmetric neutrino emission is important for building an accurate science case for the detection of CCSNe in Einstein Telescope. 

For a GW detector on the moon, SFHo\_rr can be detected out to $\sim 3$\,Mpc, SFHo\_nr out to $\sim 600$\,kpc, SFHx\_rr out to $\sim 2$\,Mpc, and SFHx\_nr out to $\sim 600$\,kpc. Due to the better low-frequency sensitivity on the moon, there is a much larger dependence on source orientation, with the detection distances for SFHx\_rr differing by over 1\,Mpc for different source orientations, which shows that including multiple source orientations is important for detectability studies for CCSNe in low-frequency GW detectors. This is certainly an underestimation of the actual maximum distance that we could detect CCSNe on the moon. This is because our waveforms are too short in duration, and can only go down to frequencies of 2\,Hz, when the full frequency band of a moon detector goes down to $\sim 0.1$\,Hz. The amplitude at low frequencies would also grow further with a longer simulation time. 

\begin{figure}
\includegraphics[width=\columnwidth]{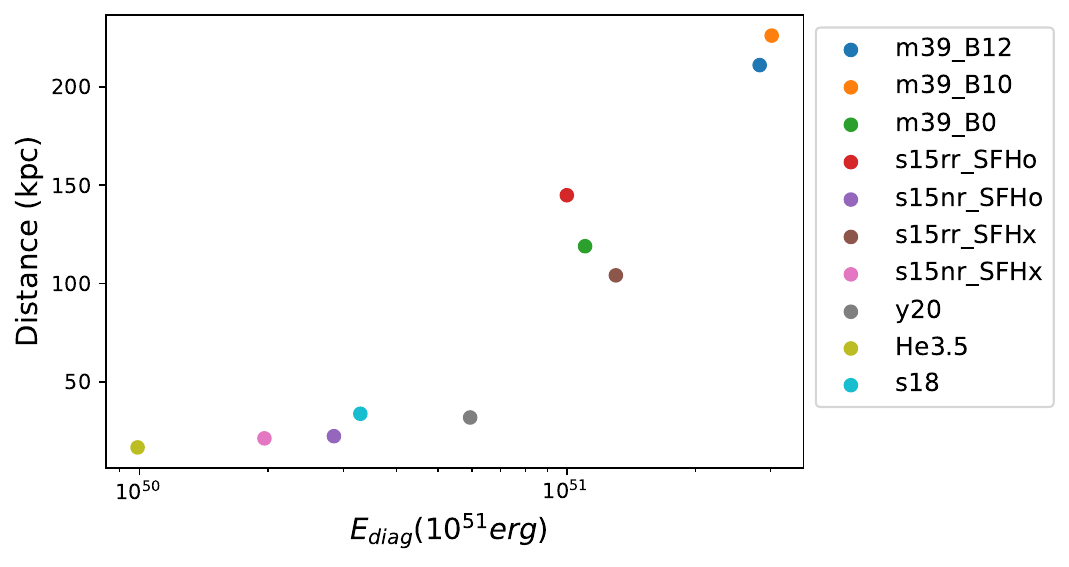}
\caption{Maximum detection distance versus diagnostic explosion energy for our new models 
along with models from our previous studies \citep{2019MNRAS.487.1178P, 2020MNRAS.494.4665P, 2023MNRAS.522.6070P}.
The maximum detection distance for the models in this work are similar to those from previous neutrino-driven explosions simulated without magnetic fields.
 }
\label{fig:gw_distance}
\end{figure}

In Figure~\ref{fig:gw_distance}, we compare the detection distances for our new models in this paper to our models from our previous studies. We include a non-rotating low mass model he3.5 \citep{2019MNRAS.487.1178P}, two typical non-rotating neutrino-driven explosions models s18 and y20 \citep{2019MNRAS.487.1178P, 2020MNRAS.494.4665P}, a rapidly rotating model with no magnetic fields model m39 \citep{2020MNRAS.494.4665P} and two models m39\_B12 and m39\_B10 both with rapid rotation and strong magnetic fields \citep{2023MNRAS.522.6070P}. We show the maximum detection distance in a single Advanced LIGO detector. There is a clear relationship between the explosion energy of the source and the GW detectability. Our new models have comparable detection distances to previous models without magnetic fields. Typical CCSNe have explosion energies of $10^{51}$\,erg, but that energy is not reproduced in the majority of 3D CCSN simulations. As our rotating GW signals have energies closer to $10^{51}$\,erg, they may be more representative of the actual maximum detection distance for typical CCSNe in Advanced LIGO. A network of GW detectors, and more complete waveforms, would further enhance the detection distances for CCSNe.  

\begin{table*}
\centering
 \begin{tabular}{||c c c c c c c c c c c c c||} 
 \hline
  Model Name & Theta & Phi & LIGO(m) (kpc) & ET(m) (kpc) & ET(m+n) (kpc) & Moon(m+n) (kpc) & Frequency (Hz) \\ 
 \hline\hline
  SFHo\_rr & 0 & 0 & 136.4 & 1417.3 & 1760.6 & 3217.0 & 444 \\
 \hline
   & 180 & 0 & 128.1 & 1340.2 & 1401.3 & 3069.0 & 712 \\ 
 \hline
  & 90 & 0 & 136.7 & 1405.7 & 1467.0 & 3318.9 & 1115 \\
 \hline
  & 90 & 90 & 144.9 & 1508.3 & 1569.6 & 3543.1 & 680 \\
 \hline
  & 90 & 180 & 135.2 & 1402.3 & 1445.2 & 3179.8 & 999 \\
 \hline
  & 90 & 270 & 135.3 & 1409.2 & 1471.3 & 3321.2 & 793 \\
 \hline \hline
  SFHo\_nr & 0 & 0 & 17.8 & 171.9 & 185.8 & 468.2 & 1057 \\
 \hline
  & 180 & 0 & 17.9 & 173.0 & 231.1 & 602.9 & 1208 \\
 \hline
  & 90 & 0 & 21.7 & 210.2 & 267.3 & 742.8 & 1396 \\
 \hline
  & 90 & 90 & 22.5 & 219.7 & 259.9 & 692.5 & 1248 \\
 \hline
  & 90 & 180 & 21.8 & 211.8 & 274.4 & 686.4 & 1264 \\
 \hline
  & 90 & 270 & 23.0 & 226.1 & 260.1 & 703.7 & 970 \\
 \hline \hline 
  SFHx\_rr & 0 & 0 & 83.0 & 862.6 & 937.6 & 937.6 & 1472 \\
 \hline
  & 180 & 0 & 90.7 & 947.6 & 962.8 & 2098.8 & 1472 \\
 \hline
  & 90 & 0 & 108.3 & 1125.0 & 1218.5 & 2835.1 & 1437 \\
 \hline
  & 90 & 90 & 105.7 & 1107.7 & 1226.0 & 2874.9 & 1570 \\
 \hline
  & 90 & 180 & 110.3 & 1159.4 & 1234.7 & 2760.6 & 1521 \\
 \hline
  & 90 & 270 & 108.9 & 1140.3 & 1252.3 & 2879.6 & 1640 \\
 \hline \hline
  SFHx\_nr & 0 & 0 & 19.2 & 190.7 & 233.9 & 608.2 & 1118 \\
 \hline
  & 180 & 0 & 16.2 & 160.4 & 239.5 & 633.9 & 1146 \\
 \hline
  & 90 & 0 & 17.8 & 179.8 & 211.2 & 560.6 & 1079 \\
 \hline
  & 90 & 90 & 21.4 & 220.7 & 259.5 & 680.7 & 876 \\
 \hline
  & 90 & 180 & 20.5 & 207.0 & 234.7 & 613.6 & 1149 \\
 \hline
  & 90 & 270 & 22.6 & 229.6 & 265.4 & 700.5 & 1153 \\
 \hline 
\end{tabular}
\caption{For each of our four simulations we show the model name, the six different source angles used, the maximum detection distances in LIGO for the matter component only, the Einstein Telescope for matter only (m) and matter plus neutrinos (m+n), the moon detector for both matter and neutrinos. 
For each source orientation, we also show the frequency where the GW energy is highest.
We define the maximum detectable distance as the distance required for an optimal SNR of 8.}
\label{tab:detection}
\end{table*}

\section{Signal interpretation and universal relations}
\label{sec:universal} 

\begin{figure*}
\includegraphics[width=\columnwidth]{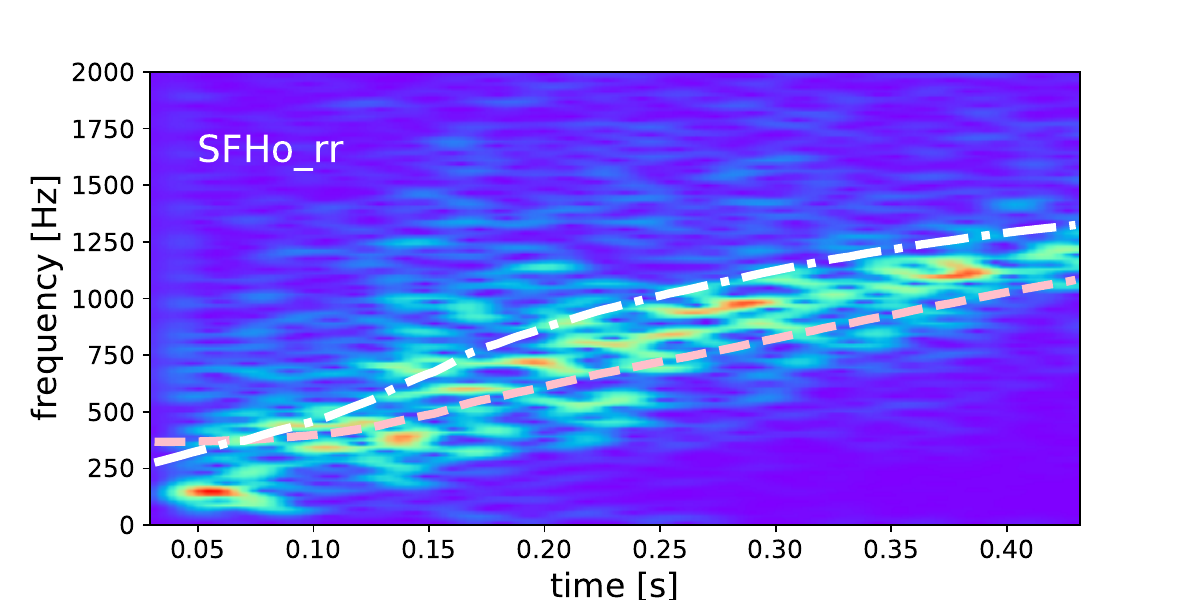}
\includegraphics[width=\columnwidth]{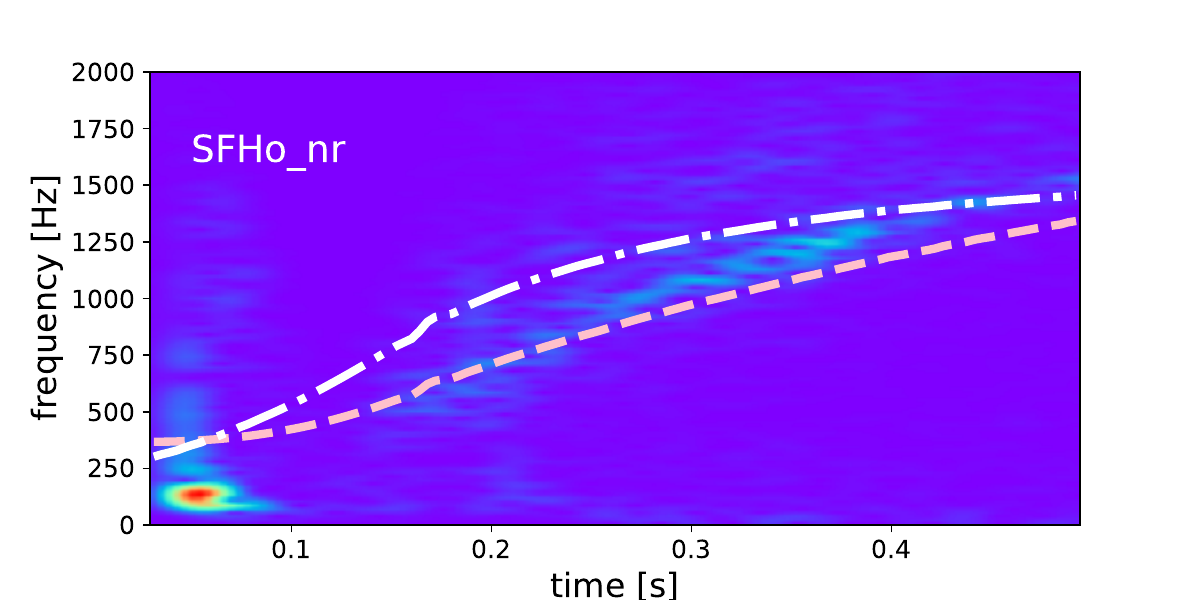}
\includegraphics[width=\columnwidth]{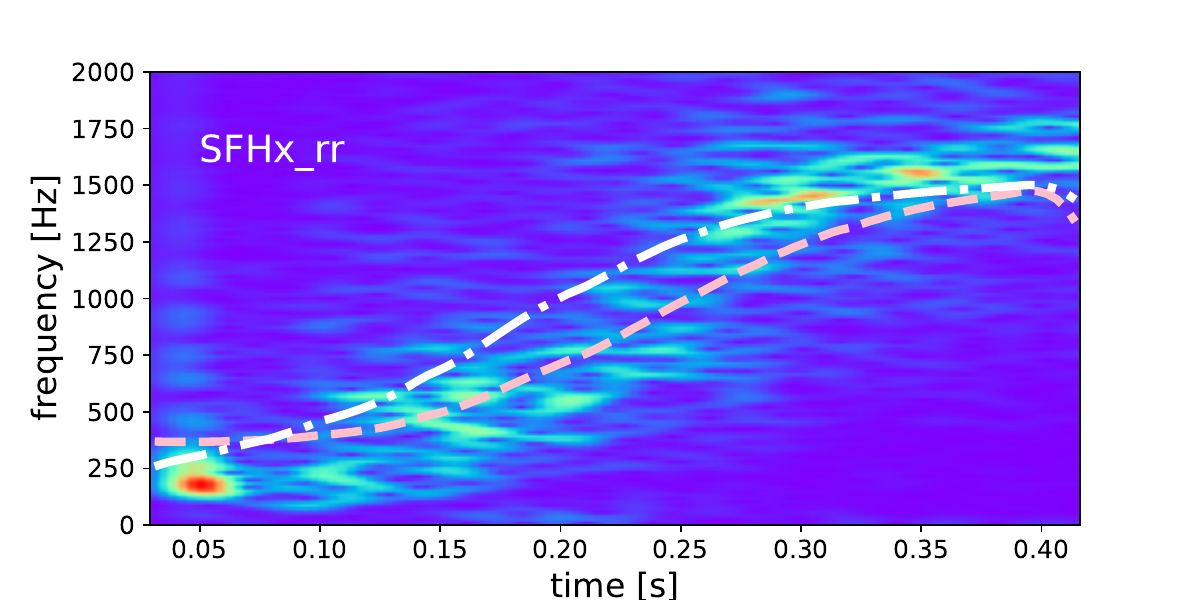}
\includegraphics[width=\columnwidth]{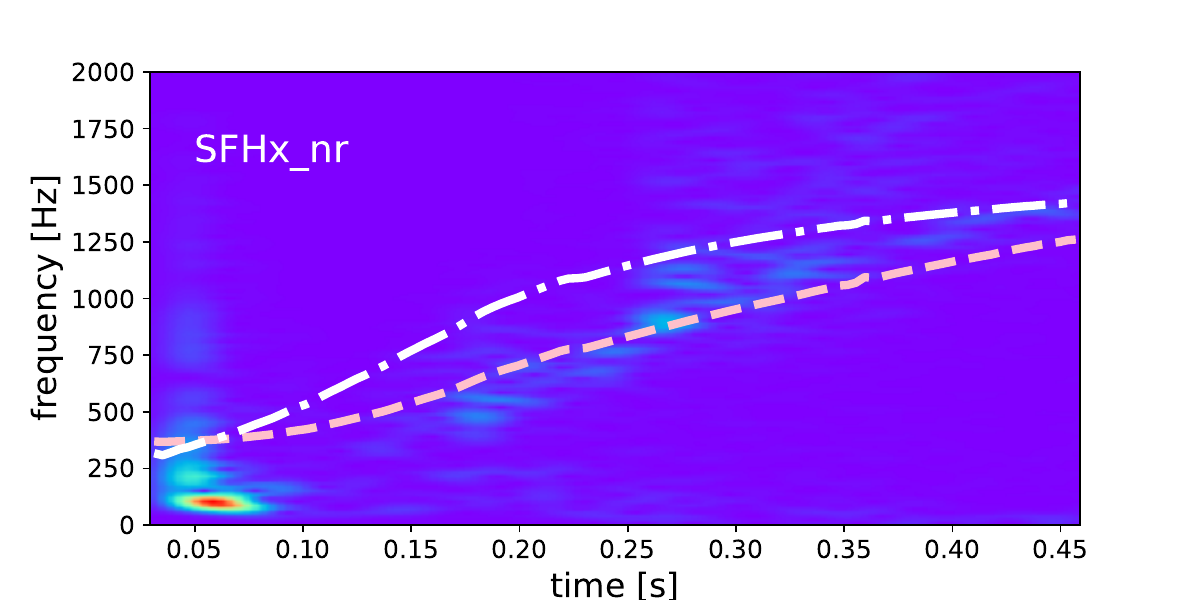}
\caption{The spectrograms of our GW signals, as shown in Figure \ref{fig:spectrograms}, with our models fit to the universal relation from \citet{torres-forne_universal_2019} in white, and \citet{2021PhRvD.104l3009S} in pink. Both relations are a reasonably good fit for our models, even when the source is rapidly rotating. Top left is SFHo\_rr, top right is SFHo\_nr, bottom left is SFHx\_rr, and bottom right is SFHx\_nr. }
\label{fig:universal}
\end{figure*}

Many groups have now formulated universal relations, that are designed to formulate the relationship between the GW frequency of the modes and properties like the PNS mass, PNS radius and the shock radius for every CCSN GW signal \citep{torres-forne_universal_2019, 2021PhRvD.104l3009S}. These universal relations have been used in GW studies for the purpose of GW signal interpretation \citep{2022PhRvD.105f3018P, 2023arXiv230110019B}. However, the current universal relations do not account for rotation or magnetic fields. They also do not account for the different possible physical 
mechanisms responsible for the GWs \citep{2023PhRvD.107d3008M}, which we have not investigated in this study. 

To see how well our models fit the current proposed universal relations, we plot the predicted frequency for the $2g^{2}$ mode from \citet{torres-forne_universal_2019}, and we also try the universal relation from \citet{2021PhRvD.104l3009S}.
The universal relation from \citet{2021PhRvD.104l3009S} is given by, 
\begin{equation}
    f/\mathrm{kHz} = -1.410 - 0.443\ln{x} + 9.337x -6.714x^{2}
\end{equation}
where $x$ is given by
\begin{equation}
    x = \left( \frac{M_\mathrm{PNS}}{1.4M_{\odot}} \right)^{1/2}  \left( \frac{R_\mathrm{PNS}}{10\,\mathrm{km}} \right)^{-3/2} , 
\end{equation}
where $M_\mathrm{PNS}$ is the PNS mass, and $R_\mathrm{PNS}$ is the PNS radius. The universal relation for the $2g^{2}$ mode from \citet{torres-forne_universal_2019} is given by the form,
\begin{equation}
f = 5.88x - 86.2x^{2} + 4.67x^3
\end{equation}
where $x$ is given by 
\begin{equation}
x = \frac{ M_\mathrm{PNS}}{R^{2}_\mathrm{PNS}}.
\end{equation}

Our fit to the universal relations is shown in Figure \ref{fig:universal}, where we defined the PNS radius as the radius where the density is $10^{11}\,\mathrm{g}\,\mathrm{cm}^{-3}$. 
We find that both of the universal relations provide a reasonable fit to our models, even when the model is rapidly rotating, although this is more difficult to see in the non-rotating spectrograms, as they are dominated by the 2D prompt convection. The universal relation from \citet{torres-forne_universal_2019} predicts a higher frequency than the relation from \citet{2021PhRvD.104l3009S}. 

Note, however, that the frequencies computed in the pseudo-Newtonian approximation used in our simulations will be systematically too high. The relations 
may thus provide a good description of the mode trajectory, perhaps after additional calibration. 
Our results show that some caution should be used when applying universal relations for the interpretation of the source properties from a real CCSN GW detection.

To be able to use the universal relations to determine the source parameters of a CCSN, we need to be able to reconstruct the signal from the GW data. Current searches for GWs from CCSNe show that a high SNR of $\sim 20$ is needed before the signal can be detected, and waveform reconstruction methods have needed a SNR of $\sim 20$ for the reconstructed waveform to have a 50\% overlap with the original waveform \citep{Szczepanczyk_2021, 2022PhRvD.106f3014R}.
Therefore, there also needs to be improvements to CCSNe GW searches and waveform reconstruction techniques for us to be able to confidently estimate the parameters of the source when a real CCSN GW signal occurs.

\section{Conclusions}
\label{sec:conclusion}

In recent years, there has been a significant increase in the number of long-duration 3D simulations of GWs from typical CCSNe with no rotation and no magnetic fields. The GW emission for neutrino-driven explosions still has some uncertainties, e.g., regarding signal amplitude, but the typical overall shape of the signal in the time-frequency space is well understood. However, there has only been a small number of GW signals from 3D simulations that include magnetic fields and rotation that are continued well beyond the core-bounce phase. Therefore, in this paper, we performed four simulations of the explosions of a $15\,M_{\odot}$ star with weak pre-collapse magnetic fields to determine the impact of the magnetic fields on the GW signal. We included both rotating and non-rotating models and used two different EoS.  

The end times of our simulations are between 0.42\,s and 0.48\,s after core bounce, which is long enough to capture the main f/g-mode and any GW emission due to the SASI. We gave a brief discussion of the hydrodynamics that are relevant for understanding the GW emission. The rotating models undergo shock revival at $\sim 100$\,ms post bounce, and the non-rotating models undergo shock revival at $\sim 200$\,ms post bounce. The observed explosion energies are similar to those observed in models with no magnetic fields, showing that our weak magnetic fields did not have a big impact on the explosion dynamics. The rotating models quickly form unstable jets. 

We also analysed the properties of the PNS, as the GW frequency of the main f/g-mode is closely correlated with the PNS properties. We observe smaller PNS masses in rotating models, which is consistent with previous work. We have one model SFHx\_rr that has a smaller PNS radius than all the other models. This results in the GW emission for that model reaching higher frequencies than the other three models. 

The GW amplitudes, and maximum detection distances are consistent with our previous models that do not include magnetic fields. Stronger magnetic fields would be needed for a CCSN to produce powerful GW amplitudes from a magnetorotational explosion. The GW frequencies are higher than in our previous neutrino-driven explosions, which is likely due to the simulation set up, but may also be partly due to the presence of the magnetic fields. 

We also investigate the low-frequency GW emission due to both the asymmetric matter and the asymmetric emission of neutrinos. The low-frequency GW emission from matter starts to grow to tens of cm in amplitude before the end of the simulation time, and would likely have reached a larger amplitude if we were able to continue the simulations for a longer duration. However, the GWs due to the asymmetric emission of neutrinos have very high amplitudes, a few hundred cm at low-frequencies, by the end of the simulation time. 

We investigated the impact of the low-frequency GW emission on the detectability of CCSNe, as this aspect of the GW signal is neglected in most detectability studies. We show that the GW amplitude at low-frequencies can vary significantly at different source angles. The GWs from the asymmetric emission of neutrinos has no impact on the detectability of CCSNe in Advanced LIGO, as it occurs below the frequency band of current GW detectors. However, we find that including this aspect of the GW emission can have a significant impact on the detectability of CCSNe in the next generation detector Einstein Telescope, which shows that it is important that future studies include this feature of the GW emission when developing the science case for detecting CCSNe in next generation detectors. We also show that CCSNe are a promising source for a moon-based GW detector, even if we cannot predict our GW frequencies low enough to cover the entire moon detection band, due to the short duration of our simulations.

\section*{Acknowledgements}

Authors JP and BM are supported by the Australian Research Council (ARC) Centre of Excellence (CoE) for Gravitational Wave Discovery (OzGrav) project numbers CE170100004 (JP, BM) and CE230100016 (JP). JP is supported by the ARC Discovery Early Career Researcher Award (DECRA) project number DE210101050. BM acknowledges supported by the ARC through Future Fellowship FT160100035 and Discovery Project DP240101786. We acknowledge computer time allocations from Astronomy Australia Limited's ASTAC scheme, the National Computational Merit Allocation Scheme (NCMAS), and from an Australasian Leadership Computing Grant. Some of this work was performed on the Gadi supercomputer with the assistance of resources and services from the National Computational Infrastructure (NCI), which is supported by the Australian Government, and through support by an Australasian Leadership Computing Grant. Some of this work was performed on the OzSTAR national facility at Swinburne University of Technology. OzSTAR is funded by Swinburne University of Technology and the National Collaborative Research Infrastructure Strategy (NCRIS).

\section*{Data Availability}

The data from our simulations will be made available upon reasonable requests made to the authors.



\bibliographystyle{mnras}
\bibliography{example} 



%
%
%


\bsp	
\label{lastpage}
\end{document}